\documentclass[fleqn,usenatbib]{mnras}
\usepackage[utf8]{inputenc}
\usepackage{mathrsfs}
\usepackage{natbib}
\usepackage{amsmath}
\usepackage{amsfonts}
\usepackage{amssymb}
\usepackage{graphicx}
\usepackage{hyperref}
\usepackage[usenames,dvipsnames]{color}
\usepackage{mathrsfs}
\usepackage{xcolor}

\makeatletter

\providecommand{\tabularnewline}{\\}

\usepackage{mathrsfs}

\newcommand{\abs}[1]{\left\vert#1\right\vert}
\newcommand{\set}[1]{\left\{#1\right\}}

\newcommand{\eps}{\varepsilon}

\newcommand{\ord}[1]{\textrm{ord}\left(#1\right)}
\newcommand{\zhat}{\mathbf{\hat{z}}}
\newcommand{\yhat}{\mathbf{\hat{y}}}
\newcommand{\xhat}{\mathbf{\hat{x}}}

\newcommand{\ham}{\mathscr{H}}

\definecolor{darkgreen}{RGB}{31, 207, 31}

\makeatother

\title[Resonant Dynamical Friction]{Resonant Dynamical Friction Around a Super-Massive Black Hole: Analytical Description}

\author[Y. B. Ginat et al.]{Yonadav Barry Ginat,$^{1}$\thanks{ginat@campus.technion.ac.il}
Taras Panamarev,$^{2,3}$
Bence Kocsis,$^{2}$
and
Hagai B. Perets$^{1,4}$
\\
$^1$ Faculty of Physics, Technion -- Israel Institute of Technology, Haifa, 3200003, Israel\\
$^2$ Rudolf Peierls Centre for Theoretical Physics, Parks Road, OX1 3PU, Oxford, UK\\
$^3$ Fesenkov Astrophysical Institute, Observatory 23, 050020 Almaty, Kazakhstan \\
$^4$ Department of Natural Sciences, The Open University of Israel, 1 University Road, Ra'anana, 4353701, Israel\\
}

\date{Accepted XXX. Received YYY; in original form ZZZ}

\pubyear{2022}
\begin{document}
\label{firstpage}
\pagerange{\pageref{firstpage}--\pageref{lastpage}}
\maketitle

\begin{abstract}
We derive an analytical model for the so-called phenomenon of `resonant dynamical friction', where a disc of stars around a super-massive black hole interacts with a massive perturber, so as to align its inclination with the disc's orientation. We show that it stems from a singular behaviour of the orbit-averaged equations of motion, which leads to a rapid alignment of the argument of the ascending node $\Omega$ of each of the disc stars, with that of the perturber, $\Omega_{\rm p}$, with a phase-difference of $90^\circ$. This phenomenon occurs for all stars whose maximum possible $\dot{\Omega}$ (maximised over all values of $\Omega$ for all the disc stars), is greater than $\dot{\Omega}_{\rm p}$; this corresponds approximately to all stars whose semi-major axes are less than twice that of the perturber. The rate at which the perturber's inclination decreases with time is proportional to its mass and is shown to be much faster than Chandrasekhar's dynamical friction. We find that the total alignment time is inversely proportional to the root of the perturber's mass.
This persists until the perturber enters the disc. The predictions of this model agree with a suite of numerical $N$-body simulations which we perform to explore this phenomenon, for a wide range of initial conditions, masses, \emph{etc.}, and are an instance of a general phenomenon. Similar effects could occur in the context of planetary systems, too.
\end{abstract}




\section{Introduction}
\label{sec:introduction}

Dynamical friction is a process whereby collective gravitation interactions between a massive body and a large number of smaller ones (whose total mass dominates) lead to a net effect on the motion of the former, in a manner mimicking friction \citep{Chandrasekhar1943,Binney}. The usual approach (e.g. \citealt{LBK1972,TremaineWeinberg1984}) to modelling dynamical friction treats the potential of the large mass as a perturbation to the whole system's Hamilton's equations of motion, which are solved perturbatively. At first order, the perturber scatters smaller objects from one orbit (in the background potential) to another, and so creates an over-density behind it; the force exerted on the perturber is then a consequence of the net gravitational pull of this wake.

Dynamical friction may also arise as a consequence of orbit-averaged interactions: suppose that the entire system consists of particles orbiting in the potential $\varphi_0$ of some object, which is so massive that it is entirely unaffected by them. Every particle primarily follows the orbits of this potential, but secular effects may appear due to their interaction with other particles \citep{Arnoldetal2006,Morbidelli2002}, which may be accounted for by averaging the correction to the potential, $\varphi_1$, due to the self-gravity of the particles, over the orbits of $\varphi_0$. If one of the particles is much heavier than the others, these orbit-averaged equations describe a dynamical-friction-like force (e.g. \citealt{TremaineWeinberg1984}). Indeed, such a phenomenon was recently observed numerically in the context of an intermediate-mass black hole (IMBH) entering a disc of stars around a super-massive black hole (SMBH) by \citet{Szolgyenetal2021}, which was termed `resonant dynamical friction'. This process pertains to the alignment of the inclination of the IMBH's orbit with the disc, and the main purpose of this paper is to offer an analytical explanation of this phenomenon. A possibly-related process occurs in rotating clusters \citep{SzolgenKocsis2018,Gruzinovetal2020}.

In the inner-most regions of the Galactic centre, there is a disc of young massive stars \citep{LevinBeloborodov2003,PaumardEtAl2006,Gallego-Canoetal2018,Alietal2020,Schoedeletal2020,Fellenbergetal2022} orbiting around the SMBH Sgr A*, whose mass is $M_\bullet \approx 4 \times 10^6~M_\odot$ \citep{Schoedeletal2002,Eisenhaueretal2005,Ghezetal2005,Ghezetal2008,GravityCollaboration2018,EventHorizonTelescope2022}. The observed distribution of the arguments of the ascending node of these stars exhibits unusual features \citep{Gallego-Canoetal2018,Alietal2020,Fellenbergetal2022}, which has been interpreted as evidence for two distinct discs (e.g. \citealt{Alietal2020}), but the existence of a second separate disc may be moot \citep{Fellenbergetal2022}. Many phenomena can lead to features in the distribution of the arguments of the ascending node, $\Omega$, for example in the context of the Solar system: evidence for the hypothesised existence of Planet 9 is related, in part, to a non-uniform distribution of $\Omega$ (see, e.g. \citealt{BatyginBrown2016}).
In this paper, we show that a massive perturber aligns the stars' arguments of ascending node with a $90^\circ$ offset with respect to its own, and this configuration in turn causes resonant dynamical friction driving the alignment of the perturber's orbital inclination.

Numerical simulations by \citet{Szolgyenetal2021} showed that an IMBH of mass $m_{\rm p} = \textrm{a few} \times 1000~M_\odot$ on an inclined orbit with respect to a stellar disc around an SMBH, warps the disc and the relative inclination decreases until the IMBH becomes embedded in the disc. The time-scale for this process was found to be similar to the time-scale of resonant relaxation \citep{RauchTremaine1996}, but much shorter than the time-scale of inclination change due to Chandrasekhar dynamical friction.\footnote{Generally, `resonance' refers to the relation between the fundamental frequencies of motion of different bodies in the system \citep{RauchTremaine1996}. If the mean-field potential admits action-angle variables, resonance occurs between all bodies for which the angles change respectively with frequencies $(\Omega_1,\Omega_2,\Omega_3)$ such that they satisfy a resonance condition: $n_1\Omega_1+n_2\Omega_2+ n_3 \Omega_3=0$, with integer $(n_1,n_2,n_3)$ prefactors, not all zero. In particular, for vector resonant relaxation, the mean field potential is generated by the SMBH and the spherical  extended stellar mass distribution and/or general relatistic corrections, which drive elliptic motion and apsidal precession, respectively, while the argument of node is fixed, $\Omega_3=0$. The resonance condition holds with $n_1=n_2=0$ and $n_3$ arbitrary for all bodies in the system, leading to a rapid coherent change in the $z$-component of the angular momentum vectors \citep{RauchTremaine1996}. Physically, the orbits in the spherical potential cover punctured discs which interact coherently until the angular momentum vectors reorient. The coherent accumulation of torques due to this global resonance drives an accelerated evolution relative to the energy diffusion driven by stochastic incoherent two-body encounters. }
This study also compared an $N$-ring code and $N$-body simulations, and found a very good agreement between the two approaches; thus, the underlying physical mechanism is probably the same as what gives rise to resonant relaxation.

While it is unknown whether IMBHs exist in the relevant mass range in the Universe, if they do, the Galactic centre is a relatively likely place to find them \citep{Yu_Tremaine2003,GoodmanTan2004,Gualandris2009, Gualandris2010,Kocsis+2011,Kocsis+2012,McKernan+2014,ArcaSedda+2019,Naoz2020, Gravity2020}.

This paper explores the orbit-averaged gravitational interactions between $N$ stars lying initially in a flat disc and a perturber, which may be thought of as an IMBH, on an inclined orbit. We start by describing the problem and the orbit-averaged equations of motion in \S \ref{sec:set-up}, we then solve these equations in \S \ref{sec:early times} -- which are singular, when studied in the context of perturbation theory -- and derive solutions for the arguments of the ascending node of the disc stars and the perturber. We find that they align to have a phase-difference of $90^\circ$, and then show how resonant dynamical friction arises from these equations. Most of the calculations are done in appendices, and \S \ref{sec:early times} summarises their conclusions. Then, we perform a set of $N$-body simulations to test our model, which are described in \S \ref{sec:simulation shortened}, and whose results are presented in \S \ref{sec:results}. We discuss our results in \S \ref{sec:discussion} and summarise in \S \ref{sec:summary}.


\section{Set-up}
\label{sec:set-up}

Suppose that one has a thin disc of $N\gg1 $ stars of masses $m_n$ ($n\in\{1,\ldots,N\}$) of total mass $M_{\rm d} \equiv \sum_{n=1}^{N}m_n$, surrounding a SMBH whose mass is $M_\bullet$, and a particle of mass $m_{\rm p}$, the ``perturber'', is placed on an inclined orbit. We assume that the following mass-hierarchy holds
\begin{equation}\label{eq:masshierarchy}
 \max\set{m_n} \ll m_{\rm p} \ll M_{\rm d}  \ll M_\bullet
\end{equation}
and that the stars have small eccentricities and mutual inclinations, so the Hamiltonian is expanded to second order in them, but not in the eccentricity and the inclination of the perturber, which can be large. In the case we study here, the perturber moves under the influence of the \emph{entire} disc, while each of the disc stars is affected primarily by the perturber, but also by the other stars (treated as a perturbation to the former). This is different from the case of pairwise interactions (cf. \citealt{KocsisTremaine2015}), because the perturber is affected by the collective gravity of all disc stars.

Moreover, since vector resonant relaxation (VRR) is the dominant perturbation with respect to the Keplerian orbit around the central SMBH \citep{KocsisTremaine2011,KocsisTremaine2015,Fouvryetal2019}, the Hamiltonian (i.e. the disturbing function) is orbit-averaged both over the mean anomalies and over the arguments of periapsis. This is justified by the fact that the semi-major axes and eccentricities are approximately conserved as two-body relaxation (and consequently Chandrasekhar dynamical friction) and scalar resonant relaxation take place on much longer time-scales, which may therefore be safely ignored. Hence the leading order Keplerian term in the Hamiltonian $-\sum_{i=1}^{n} G m_i/(2a_i)$ is a constant which may be omitted when solving for the evolution of the inclinations and the arguments of ascending node, which specify the directions of the angular momentum vectors.

The Hamiltonian governing the system is
\begin{equation}
  \ham = \ham_{\rm p} + \ham_{\rm LL},
\end{equation}
where $\ham_{\rm LL}$ is the Laplace-Lagrange Hamiltonian given by equation (32) of \citet{KocsisTremaine2011} which describes the orbit-averaged gravitational interaction between a thin disc of stars on nearly circular and nearly co-planar Keplerian orbits and which reduces to a system of harmonic oscillators, as given explicitly in equation \eqref{eqn: ham LL definition}. This approximation is justified by our assumption that the inclinations and eccentricities of the disc stars are small, and corrections to $\ham_{\rm LL}$ are third order in the inclinations or eccentricities.
$\ham_{\rm p}$ describes the orbit-averaged interaction between the stars and the perturber \citep{KocsisTremaine2015,Roupas_2020}
\begin{equation}\label{eqn:ham p definition}
  \ham_{\rm p} = -G\sum_{n=1}^{N}\sum_{l=2}^{\infty}\frac{m_{\rm p}m_n}{\max\set{a_{\rm p},a_n}}P_{\ell}(0)^2 s_{pnl}\alpha^l_{pn} P_{\ell}(\cos\theta_{pn}),
\end{equation}
where $a_i$ are the semi-major axes, $\alpha_{pn} = \frac{\min\set{a_{\rm p},a_n}}{\max\set{a_{\rm p},a_n}}$,
$P_{\ell}$ is the $\ell$-th Legendre polynomial, the dimensionless coefficient $s_{pnl}$ is a function of $\alpha_{pn}$,\footnote{Explicitly, $s_{pn2}$ is
\begin{equation*}
    s_{pn2} \equiv \iint_0^\pi \frac{\mathrm{d}x\mathrm{d}y}{\pi^2} ~ \frac{\left[\min\set{1+e_{\rm in}\cos x,\frac{1+e_{\rm out}\cos y}{\alpha_{pn}}}\right]^3}{\left[\max\set{\alpha_{pn}(1+e_{\rm out}\cos x,1+e_{\rm in}\cos y}\right]^2},
\end{equation*}
where $e_{\rm in}$ is the eccentricity of the particle with the lesser semi-major axis, and similarly for $e_{\rm out}$.}
$e_{\rm p}$, $e_n$, and the multipole index $\ell$ only, where $s_{pn\ell}=1$ for circular orbits, and $\theta_{pn}$ is the angle between the angular momentum vector of the perturber and that of star $n$, \emph{viz.}
\begin{equation}\label{eq:cos(theta_pn)}
  \cos \theta_{pn} = \cos i_{\rm p} \cos i_n + \sin i_{\rm p} \sin i_n \cos \left(\Omega_{\rm p}  - \Omega_n \right).
\end{equation}

Since $\ham$ is already doubly-averaged over two of the three angle variables, the semi-major axes and eccentricities of all bodies are constant, and the only dynamical variables are their inclinations $\set{i_n}$ and arguments of ascending node $\set{\Omega_n}$. Up to constant coefficients, $\cos i_n$ and $\Omega_n$ are canonical momenta and position variables, respectively. 

In this paper we perform a perturbative expansion in the parameter $\eps \equiv m_{\rm p}/M_{\rm d} \ll 1$ keeping $M_{\rm d}$ fixed,
but the main result of this paper concerning the arguments of the ascending node will be valid even for $\eps = 1$, albeit for short times.
For two functions $f(\eps)$ and $g(\eps)$, we write $f(\eps) = O(g(\eps))$ if $\abs{\frac{f(\eps)}{g(\eps)}}$ becomes bounded as $\eps \to 0$, and $f(\eps) = \ord{g(\eps)}$ if $\displaystyle \lim_{\eps \to 0} \abs{\frac{f(\eps)}{g(\eps)}}$ is non-zero.

The expansion, however, does not simply amount to the na\"{i}ve one of setting $\ham_{\rm p} = \ord{\eps}$ and $\ham_{\rm LL} = \ord{1}$, because, as we shall see, the case $\eps = 0$ is singular, and therefore it calls for a special type of expansion. In fact, we will show that $\ham_{\rm p}$ \emph{dominates} the dynamics of the disc particles for most of the relevant time; in the notation of the introduction, one would have $\varphi_0 \equiv \ham_{\rm LL}$ and $\varphi_1 \equiv \ham_{\rm p}$. This sets the phenomenon of resonant dynamical friction apart from other dynamical friction phenomena (e.g. \citealt{TremaineWeinberg1984,Ostriker1999,Binney,Magorrian2021,BanikBosch2021,Desjacquesetal2022,DootsonMagorrian2022}), where usually $\varphi_1\ll\varphi_0$ is treated as a perturbation, which acts only to scatter medium particles from one orbit in $\varphi_0$ to another, or trap them around resonant orbits in $\varphi_0$ (where dynamical friction arises from the back-reaction of this scattering).
Here, quite the opposite occurs, and, in fact, if one did attempt to perform a na\"{i}ve expansion, one would have found zero dynamical friction force acting on the perturber, to $\ord{\eps^2}$.

\section{Perturbative Solution}
\label{sec:early times}

One can use Lagrange's equations of motion, which yield \citep{murray_dermott_2000}
\begin{align}
  \frac{\mathrm{d}i_{\rm p}}{\mathrm{d}t} & = -\frac{1}{\gamma_{\rm p}^2\sin i_{\rm p}}\frac{\partial \ham_{\rm p}}{\partial \Omega_{\rm p}} \\
  \frac{\mathrm{d}\Omega_{\rm p}}{\mathrm{d}t} & = \frac{1}{\gamma_{\rm p}^2\sin i_{\rm p}}\frac{\partial \ham_{\rm p}}{\partial i_{\rm p}},\label{eqn:omega p dot Lagrange equation of motion}
\end{align}
where the coefficient $\gamma_{\rm p}$ is defined as
\begin{equation}\label{eqn: gamma p definition}
  \gamma_{\rm p}^2 \equiv \mu_{\rm p} n_{\rm p} a_{\rm p}^2\sqrt{1-e_{\rm p}^2}
\end{equation}
where $\mu_{\rm p} \equiv m_{\rm p}M_\bullet/(m_{\rm p}+M_\bullet) \approx m_{\rm p}$, and the frequency $n_{\rm p}$ is $\sqrt{G(M_\bullet+m_{\rm p})/a_{\rm p}^3}$. Likewise, one may define $\gamma_n$ for any of the disc particles, by replacing the index $\textrm{p}$ by $n$.

Thanks to the mass hierarchy, Eq.~\eqref{eq:masshierarchy}, one might expect $\ham_{\rm p}$ to be truly a perturbation relative to the disc's self-interaction Hamiltonian $\ham_{\rm LL}$, governing particles $1,\ldots,N$; this is accurate, however, only if the disc is not thin. Here, though, the thinness of the disc implies that it is in fact $\ham_{\rm p}$ which dominates both the dynamics of the perturber and the disc, because $\ham_{\rm LL} \propto ~\left[\textrm{disc thickness}\right]^2$ for a thin disc. 
The explicit equations of motion are given in appendix \ref{appendix: equations of motion}, and are then solved perturbatively there. Additionally, we provide a test case for our results in appendix \ref{subsec:oscillator}, where $i_{\rm p}$ is also taken to be small. There the equations of motion may be solved analytically, and the solutions are used to verify the results here. 

We refer the readers to the appendices for details, and state the main results here:
At $\ord{\eps^0}$, the disc-perturber interaction in $\ham_{\rm p}$ induces a nodal precession of the perturber's argument of the ascending node, $\Omega_{\rm p}$ with a frequency $\nu_{\rm p} \propto n_{\rm p}\frac{M_{\rm d}}{M_\bullet}$. This is non-zero even in the limit $\eps \to 0$.

At the next order, the thinness of the disc introduces a very short time-scale, much shorter than $\nu_{\rm p}^{-1}$, over which the arguments of the ascending node of the disc stars align with $\Omega_{\rm p}$, with a phase difference of $90^\circ$. While this phenomenon appears at $\ord{\eps}$, it is much faster, and can be derived correctly only by explicitly accounting for this additional short time-scale. The short time-scale, $b_{{\rm p}n}$ is defined by
\begin{equation}\label{eqn:b_pn definition}
  b_{{\rm p}n} = \frac{3}{8}n_n\,\frac{m_{\rm p}}{M_\bullet} \frac{a_n}{\max\set{a_{\rm p},a_n}} s_{pn2}\alpha_{pn}^2\,\sin2i_{p}\frac{\cos^2 i_n}{\sin i_{n}}.
\end{equation}
The equation of motion for $\Omega_n$ is
\begin{equation}\label{eqn: omega n dot main text}
  \dot{\Omega}_n = \frac{1}{\gamma_n^2 }\frac{\partial \ham_{\rm p}}{\partial (\cos i_n)} = b_{{\rm p}n}\cos\left(\Omega_{\rm p}(t) - \Omega_n\right);
\end{equation}
this equation has two time-scales: $\nu_{\rm p} = \dot{\Omega}_{\rm p}$, and $b_{{\rm p}n}$. Even though $b_{{\rm p}n} \propto \eps$, it is still the case that $\abs{b_{{\rm p}n}}/\abs{\nu_{\rm p}} \gg 1$ because the disc is so thin; we show in appendix \ref{sec:alignment} that for stars with $\abs{b_{{\rm p}n}} > \abs{\nu_{\rm p}}$, the solution to this equation yields that $\Omega_n$ quickly aligns such that
\begin{equation}\label{eqn: Omega_n alignment}
  \cos(\Omega_n - \Omega_{\rm p}) = \frac{-\nu_{\rm p}}{b_{\textrm{p}n}};
\end{equation}
the alignment occurs after a time $\sim b_{{\rm p}n}^{-1} \ll \nu_{\rm p}^{-1}$, i.e. much faster than one nodal precession period.
This is a striking phenomenon: the arguments of the ascending node of the stars in the disc with $\nu_{\textrm{p}}/b_{\textrm{p}n}\leq 1$ align themselves with that of the perturber, with a $90^\circ$ phase difference!

Stars for which $\abs{\nu_{\rm p}/b_{\textrm{p}n}} > 1$ behave quite differently: for them, the relevant limit of their governing differential equation, equation \eqref{eqn:example ode limits}, is the oscillatory one, which implies that $\Omega_n(t)$ simply oscillates about $\Omega_n(0)$, regardless of $\Omega_{\rm p}$; no alignment occurs.

With these solutions for the arguments of the ascending node, one can solve the equations of motion for the inclinations, $i_n$ and $i_{\rm p}$; we do so in appendix \ref{sec: disc thickening}, and find that upon substituting the aligned $\Omega_n$ and $\Omega_{\rm p}$ the aligned stars in the disc exert a coherent torque on the perturber, which leads to a net decrease in its inclination. We show in appendix \ref{sec: disc thickening} that\footnote{Recall that $b_{{\rm p}n}$ depends on time via $i_{\rm p}$.}
\begin{equation}\label{eqn:perturber's inclination equation of motion}
    \frac{\mathrm{d}i_{\rm p}}{\mathrm{d}t} = \sum_{n: \abs{\frac{\nu_{\rm p}}{b_{\textrm{p}n}(t)}} \leq 1} \frac{b_{\textrm{p}n}(0)\cos(2i_n(0))}{\gamma_n^{-2}\cos^2(i_n(0))} \left[\frac{\cos(i_n) - \cos(i_n(0))}{\gamma_{\rm p}^2 \sin i_{\rm p}}\right].
\end{equation}
When substituting the solution \eqref{eqn:i_n of t} for $i_n(t)$, we find that at early times, $\mathrm{d}i_{\rm p}/\mathrm{d}t \propto -t / \sin i_{\rm p}$; the coefficient must, by dimensional analysis, have units of frequency squared, and we \emph{define} the dynamical friction time-scale to be the reciprocal of the root of that coefficient, i.e., we define
\begin{equation}\label{eqn:tau rdf definition}
    \frac{\mathrm{d}\cos i_{\rm p}}{\mathrm{d}t} \equiv -\frac{t}{\tau_{\rm RDF}^2},
\end{equation}
at small $t$. This definition ensures that $\tau_{\rm RDF}$ gives a prediction for the time-scale over which $i_{\rm p}$ decreases significantly. We obtain
\begin{equation}\label{eqn:tau RDF}
 \tau_{\rm RDF} \propto \frac{t_{\rm orb}}{\sin(2i_{\rm p}(0))}\frac{M_\bullet}{\sqrt{m_{\rm p}M_{\rm d}}},
\end{equation}
where
\begin{equation}\label{eqn:t orb definition}
    t_{\rm orb} \equiv \frac{2\pi}{n_{\rm p}} = \frac{2\pi a_{\rm p}^{3/2}}{\sqrt{G(M_\bullet + m_{\rm p})}}
\end{equation}
is the orbital period of the perturber and the proportionality constant is of order unity. We derive the proportionality coefficient in appendix \ref{sec: disc thickening}, but for a power-law surface density density profile $\Sigma \sim r^{-\beta}$ let us state the result here (see equation \eqref{eqn:tau RDF appendix}, and equation \eqref{eqn: tau rdf general surface density} for a general circular surface density profile):
\begin{equation}
    \begin{aligned}
    \tau_{\rm RDF} & = \frac{4}{3\pi}  \frac{1}{\sin(2i_{\rm p}(0))} \frac{M_\bullet}{\sqrt{m_{\rm p}M_{\rm d,loc}}}t_{\rm orb}\,\\ &
    \times \frac{(3-\beta)^{-1/2}}{(2-\beta)^{-1/2}}\left[\frac{2}{7-2\beta} + \frac{2}{5+2\beta}\left(1-\frac{a_{\rm p}^2}{R_{\rm d}^2}\right) \right]^{-1/2}.
\end{aligned}
\end{equation}
where we have introduced the local disc mass at $r=a_p$
\begin{equation}
M_{\rm d, loc} =  \frac{\rm{d} m}{\rm{d} \ln r} = 2\pi r^2 \Sigma = (2-\beta)\left(\frac{a_{\rm p}}{R_{\rm d}}\right)^{2-\beta}  M_d
\end{equation}
Note the mass scaling $\tau_{\rm RDF} \propto m_{\rm p}^{-1/2}$ but $di_{\rm p}/dt\propto m_{\rm p}$ (Eq.~\ref{eqn:perturber's inclination equation of motion}) as expected for dynamical friction.\footnote{These results are in agreement with the numerical results in figures 5 and 6 of \cite{Szolgyenetal2021} (cf. figure \ref{fig:alignment time scale} below).} In contrast, for Chandrasekhar dynamical friction, \citep{Szolgyenetal2021}
\begin{equation}\label{eq:CDF}
    \left(\left.\frac{\rm{d}\iota_{\rm p}}{\rm{d} t}\right|_{\rm CDF}\right)^{-1} = \frac{2\sin \iota_{\rm p} \sin^3 (\iota_{\rm p}/2)}{\ln \Lambda} \frac{M_{\bullet}^2}{m_{\rm p}M_{\rm d,loc}} t_{\rm orb}\,.
\end{equation}
This implies that the total alignment timescale for resonant dynamical friction is reduced by a factor of order $(m_{\rm p}M_{\rm d,loc})^{1/2}/M_\bullet$ in comparison.

We expect equation \eqref{eqn:perturber's inclination equation of motion} to be valid under the following sufficient conditions: $\eps \ll 1$ and the disc is thin relative to $\eps$ (i.e., the perturber has not yet entered the disc); in other words, we need
\begin{equation}\label{eqn: alignment condition}
    \sin(i_n) M_{\rm d} \ll \sin(i_{\rm p}) m_{\rm p}.
\end{equation}
We expect, however, that it will remain approximately accurate until the perturber enters the disc. The reason for that is, that the time-scale for the disc's thickening (cf. equation \eqref{eqn:i_n of t}) is comparable to $\tau_{\rm RDF}$. The node alignment takes place on a time-scale $b_{{\rm p}n}(0)^{-1}$, which is shorter than $\tau_{\rm RDF}$ by a factor of $\sim \sin[i_n(0)]/\sqrt{\eps}$.\footnote{In this case, our having defined $\tau_{\rm RDF}$ in the early time limit does not pose a problem. The coefficient of $t$ changes with time as more and more stars fall out of the nodal alignment, by equation \eqref{eqn:perturber's inclination equation of motion}. Hence, the total time until the perturber enters the disc is just the inverse of the right-hand-side of equation \eqref{eqn:tau rdf definition}, where $t$ is evaluated as the time when a considerable fraction of the disc stars stop being aligned -- which is just, again, $\tau_{\rm RDF}$.}

\section{Numerical Simulations}
\label{sec:simulation shortened}
To test these results, we perform a set of numerical simulations, which we now describe.
We use a modified version of the direct $N$-body code phi-GRAPE \citep{Harfst2007} which benefits from the 4th order Hermite integration scheme using block time-steps. Despite the original design for the GRAPE cards the code is adapted for modern GPUs and is widely used in various astrophysical simulations (see e.g. \citealt{Shuo2019, Shukirgaliyev2021}).

\begin{figure}
  \centering
  \includegraphics[width=0.48\textwidth]{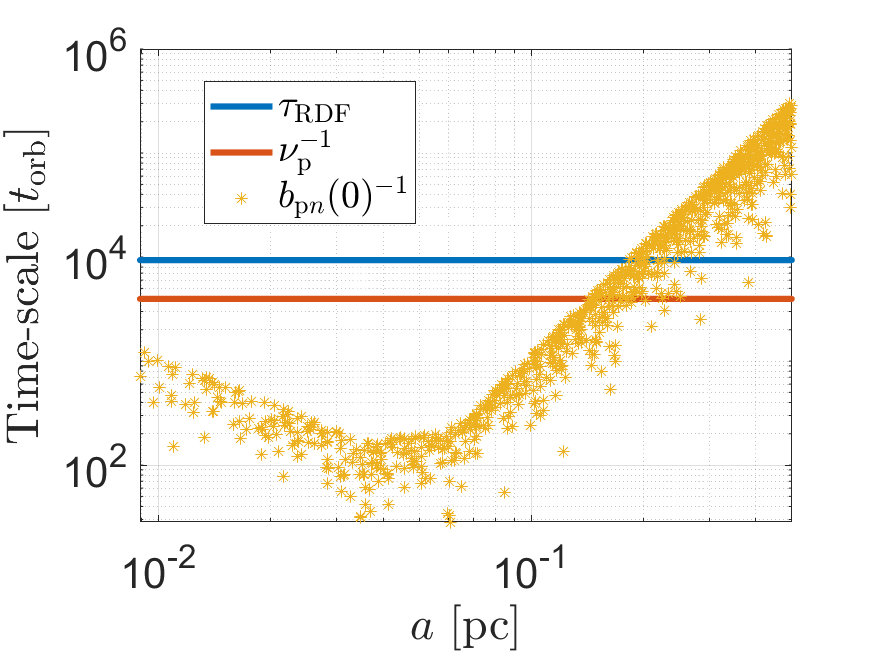}
  \caption{The relevant times-scales associated with the problem, in units of the orbital time of the perturber, $t_{\rm orb} = 524$ yr, versus the stars' semi-major axes, plotted for the same parameters as in figure \ref{fig:inclination histogram}, i.e. for the first row of table \ref{tab:runs}. The time-scales are: the nodal alignment time-scale, $b_{{\rm p}n}(0)^{-1}$ (equation \eqref{eqn:b_pn definition}), in yellow asterisks, the nodal precession time-scale $\nu_{\rm p}^{-1}$ in orange, and the inclination change time-scale, $\tau_{\rm RDF}$ (equation \eqref{eqn:tau RDF}), in blue.}\label{fig:time-scales}
\end{figure}

The simulations performed here fall into two categories: those with a disc and a perturber around the SMBH only, which are closer to the analytical model described above, and some more realistic ones which also include a `live' spherical component. In addition, we also vary the initial eccentricity and inclination distributions of the disc, as well as the initial inclination of the perturber, and its mass (from $\eps = 1/8$ to $\eps = 1$). Details of the numerical set-up are described in appendix \ref{sec:simulation}; here, it suffices to show in figure \ref{fig:time-scales} a plot of the various time-scales involved in the process, for the numerical set-up we consider. The parameters specified in appendix \ref{sec:simulation} and table \ref{tab:runs} imply that in figure \ref{fig:time-scales} $t_{\rm orb} = 524$ years, and $\tau_{\rm RDF} = 9347t_{\rm orb} \approx 4.9\times 10^6$ years. The expected synchronisation time-scale of the nodes is, on the other hand, much smaller than this $\tau_{\rm RDF}$, as shown in figure \ref{fig:time-scales}, for all stars that satisfy $\nu_{\rm p} < b_{{\rm p}n}$ initially.

\section{Results}
\label{sec:results}

\begin{figure*}
    \centering
    \includegraphics[width=\linewidth]{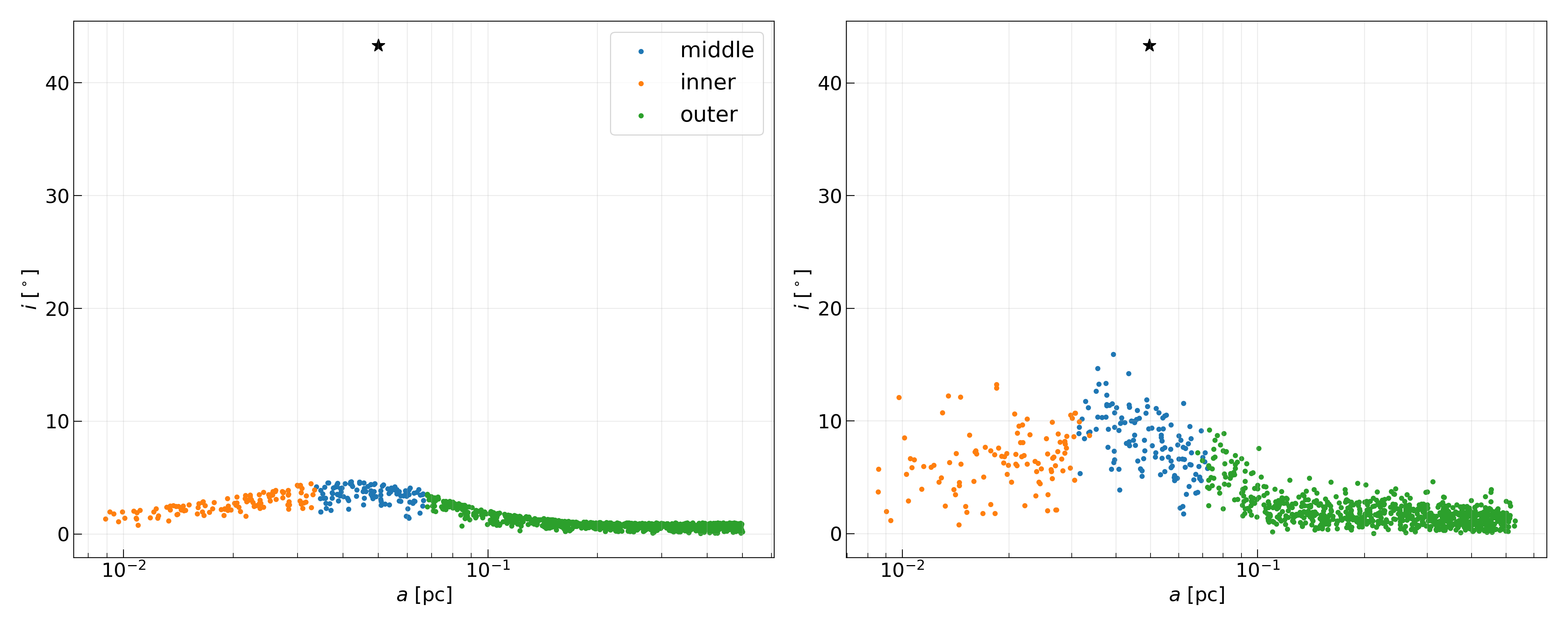}
    \caption{A scatter-plot of the inclinations of the stars as a function of their semi-major axis, after $1$ Myr. The SMBH mass is $4\times 10^6~M_\odot$, the total disc mass is $2000~M_\odot$, the perturber's mass is $250~M_\odot$, and we add a Plummer model sphere of mass $2\times 10^5~M_\odot$ to the numerical simulation to regularise the precession of the argument of pericentre. Left: the analytical prediction of equation \eqref{eqn:i_n of t}. Right: the result from the simulation with $N = 999$, corresponding to the first row of table \ref{tab:runs}. These plots show the state of the system after $1~\textrm{Myr}$.}
    \label{fig:inclination histogram}
\end{figure*}

\begin{figure*}
    \centering
    \includegraphics[width=\linewidth]{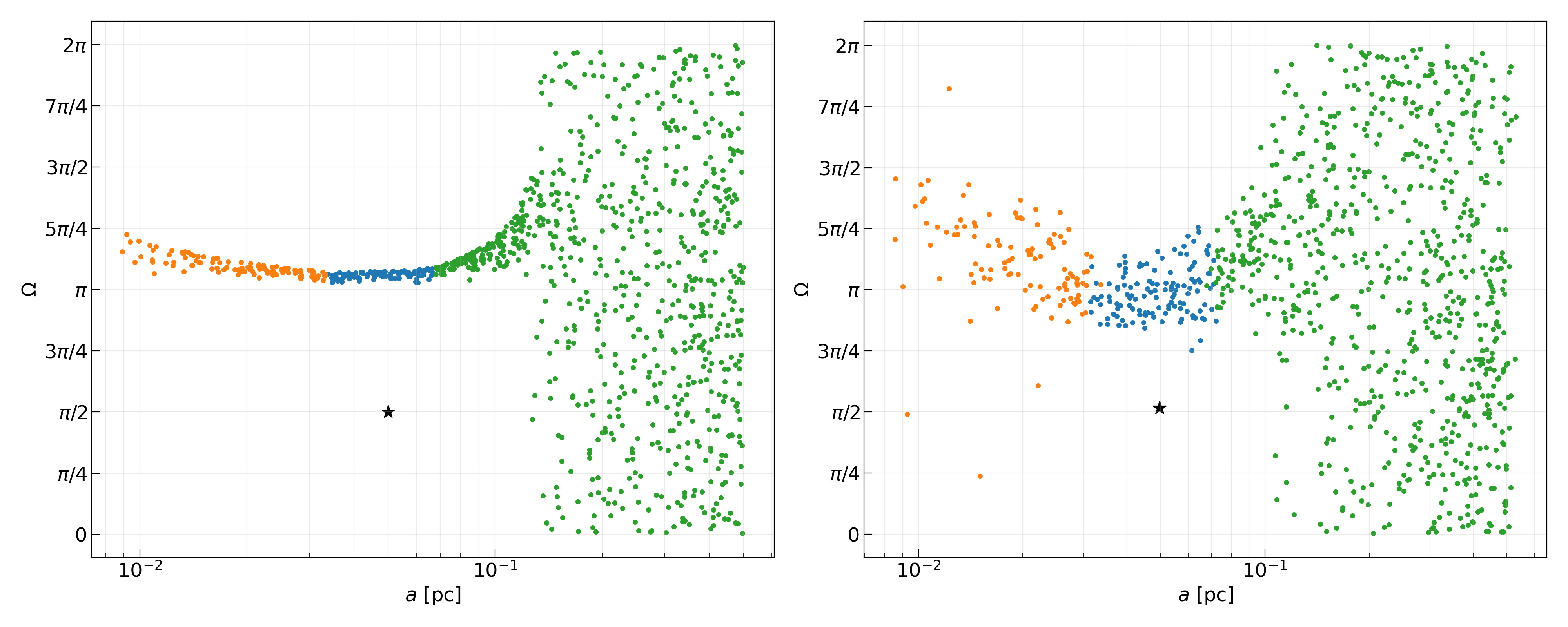}
    \caption{Like figure \ref{fig:inclination histogram}, but for the arguments of the ascending node. Left: the analytical prediction of equation \eqref{eqn: Omega_n alignment}. Right: result from the simulation.}
    \label{fig:omega histogram}
\end{figure*}

\begin{figure*}
    \centering
    \includegraphics[width=0.48\textwidth]{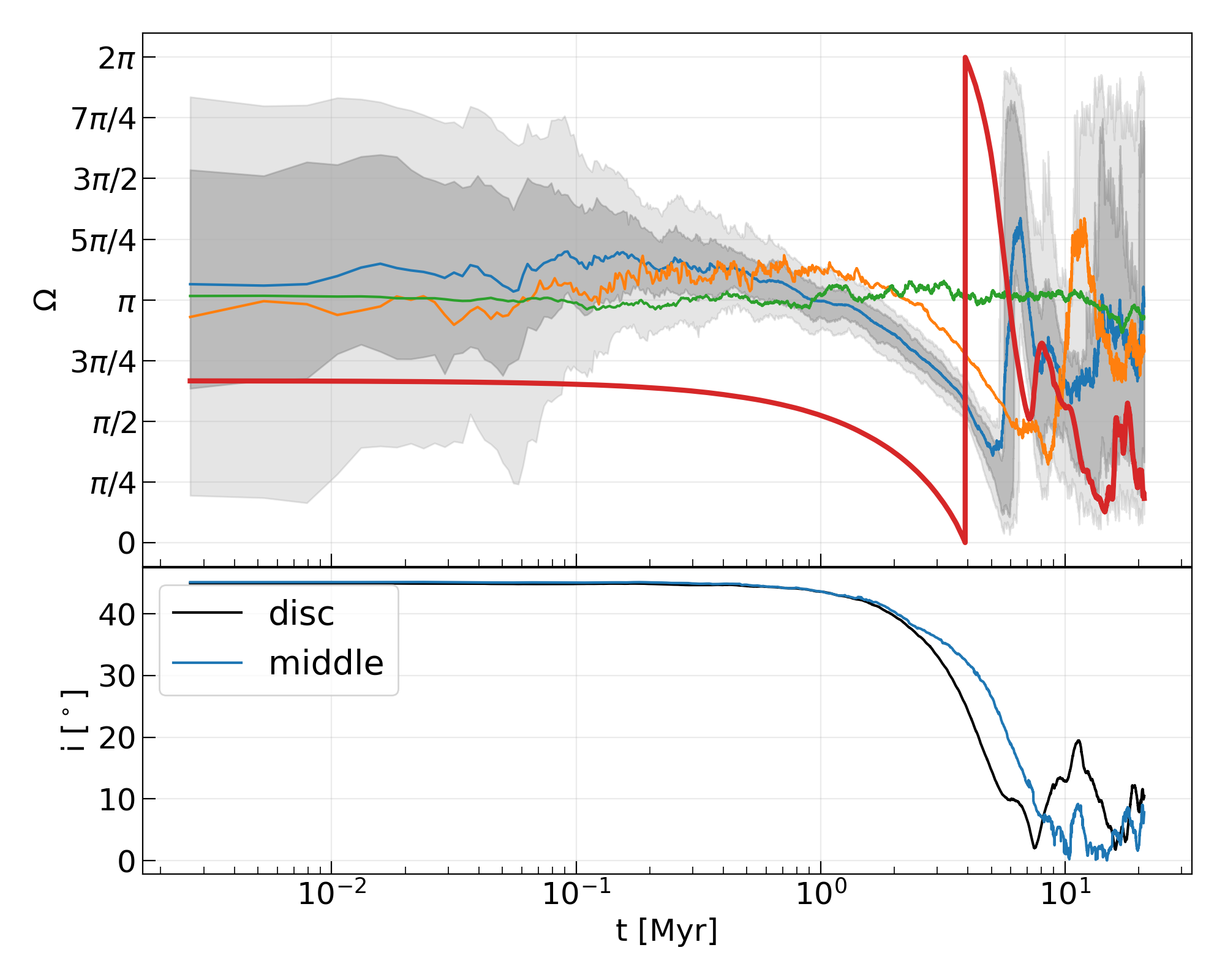}
    \includegraphics[width=0.48\textwidth]{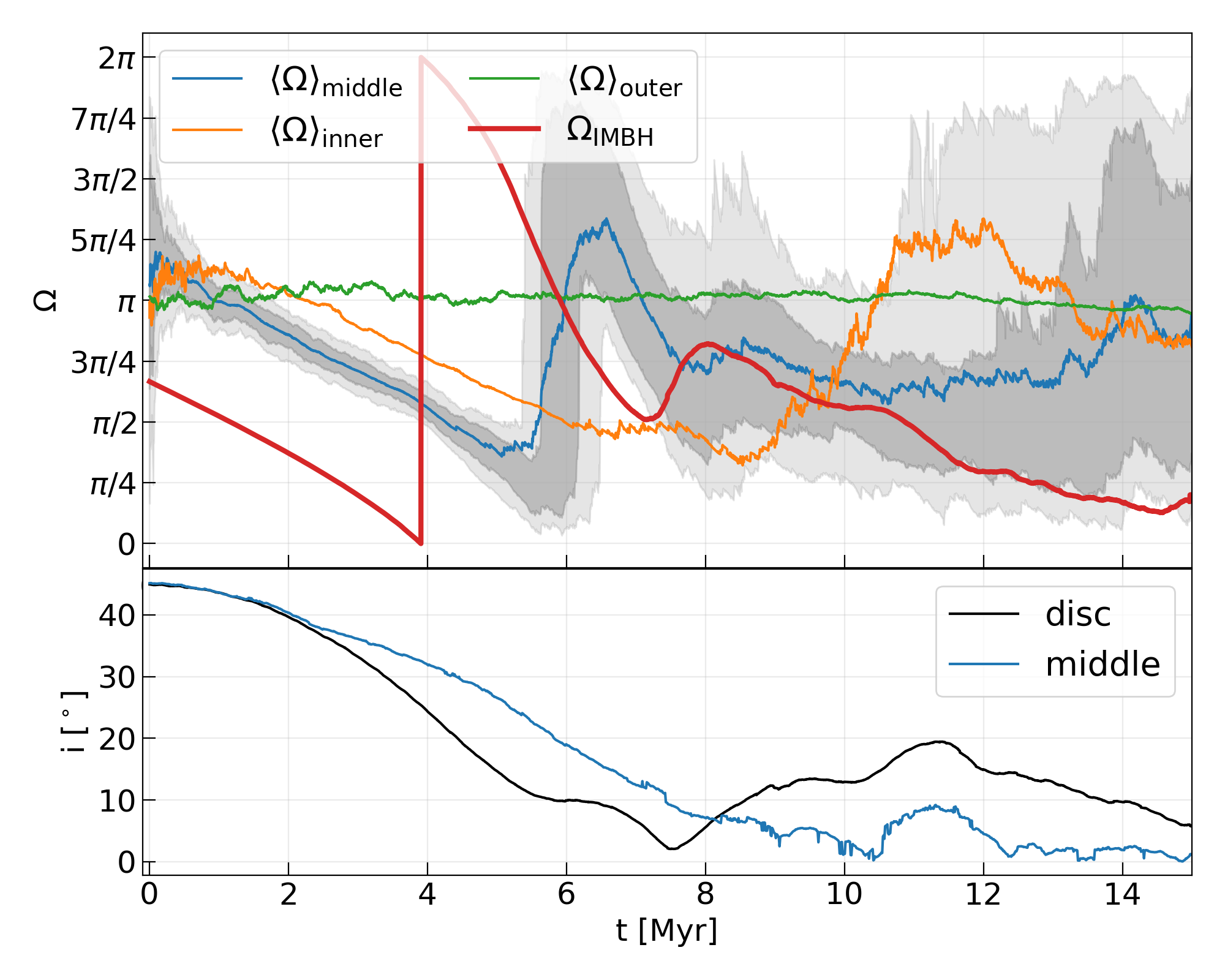}
    \caption{Top panels: time evolution of the longitudes of the ascending nodes for the IMBH and stars in the disc. The thick red line shows $\Omega$ for the IMBH, the blue line shows mean $\Omega$ of the middle stars (stars with overlapping orbits with the IMBH), the shaded region is the area between 25 and 75\% quantiles for $\left<\Omega\right>$ of the middle stars, the lightly-shaded region is the area between 10 and 90\% quantiles of the same stars. The orange and green lines show the mean $\Omega$ for the inner and outer stars, respectively. Bottom panels: the time evolution of inclination angles of the IMBH with respect to the middle stars (blue) and the whole stellar disc (black). Right: the same, but in linear scale. }
    \label{fig:omega evolution}
\end{figure*}

As one expects the predictions of \S \ref{sec:early times} to be valid until the perturber enters the disc, let us start by comparing them with those of the numerical simulations described above, when the perturber is still far from the disc, starting with those on the first row of table \ref{tab:runs} -- a thin, circular disc, with $\eps = 1/8$. A scatter-plot of the inclinations of all the stars in the disc, versus their semi-major axes, is shown in figure \ref{fig:inclination histogram} -- once derived from the analytical model of \S \ref{sec:early times}, and once from the numerical simulations. Similarly, we show a similar plot of the argument of the ascending node in figure \ref{fig:omega histogram}. Following \citet{Szolgyenetal2021}, we split the stars to 3 regions depending on the peri- and apocentres of the stars ($r_{p,*}$, $r_{a,*}$) with respect to the perturber ($r_{p,\textrm{p}}$, $r_{a,\textrm{p}}$): the inner region where the stellar orbits are within the pericentre of the perturber ($r_{a,*} < r_{p,\textrm{p}}$), the middle region where orbits of stars overlap with the perturber's orbit ($r_{p,*} \le r_{a,\textrm{p}}$ and $r_{a,*} \ge r_{p,\textrm{p}}$) and the outer region where the stellar orbits lie outside of the apocentre of the perturber ($r_{p,*} > r_{a,\textrm{p}}$).
For the readers' convenience, we also show a plot of the evolution of the average argument of ascending node of the disc within the regions as a function of time, compared with the inclination in figure \ref{fig:omega evolution}. Indeed, figure \ref{fig:omega evolution} shows that the analytical predictions of $\dot{\Omega}_{\rm p} = \textrm{const}$ and $\Omega_n - \Omega_{\rm p} \approx \pi/2$ are satisfied by the numerical simulation not only at $t=1~\textrm{Myr}$, but up to about $4.5~\textrm{Myr}$, when that they persist until the perturber enters the disc.\footnote{Or, equivalently, until the disc is broken up by it -- a situation which doesn't occur for these models, but cf. figure \ref{fig:omega t all}.}

Both the model of \S \ref{sec:early times} and the simulation exhibit a sharp transition between an alignment of $\Omega_n$ with $\Omega_{\rm p} + \pi/2$, and an essentially uniform distribution of $\Omega_n$. According to \S \ref{sec:early times} and appendix \ref{sec:alignment}, the transition occurs when $\abs{\nu_{\rm p}} = \abs{b_{\textrm{p}n}}$, i.e. when (recall that $M_\bullet \gg m_{\rm p}, m_n$ for all $n$)
\begin{equation}
    \sum_{m=1}^N \frac{m_m s_{pm2}\alpha_{pm}^2}{\max\set{a_p,a_m}} = \sqrt{\frac{a_{\rm p}}{a_n}}\frac{m_{\rm p}s_{pn2}\alpha_{pn}^2}{\max\set{a_{\rm p},a_n}} \frac{\sin i_{\rm p} \cos^2i_n}{\sin i_n}.
\end{equation}
As $\nu_{\rm p} = \ord{1}$, and $\frac{m_{\rm p}}{M_{\rm d} \sin i_n} \gg 1$, the transition can only occur at an $a_n > a_{\rm p}$, as is evident from figure \ref{fig:omega histogram}. 

A main difference in figures \ref{fig:inclination histogram} and \ref{fig:omega histogram} between the analytical model and the simulations -- the larger spread in values of orbital inclinations and longitudes of the ascending node -- is caused by effects which are neglected in the analytical model. These may include two-body interactions or scalar resonant relaxation, but a detailed investigation of these is beyond the scope of this work.

One can see that the analytical treatment both captures the essential features of the simulations, and that the simulations do indeed exhibit an alignment of the nodes, as predicted by equation \eqref{eqn: Omega_n alignment}.

\begin{figure}
  \centering
  \includegraphics[width=0.48\textwidth]{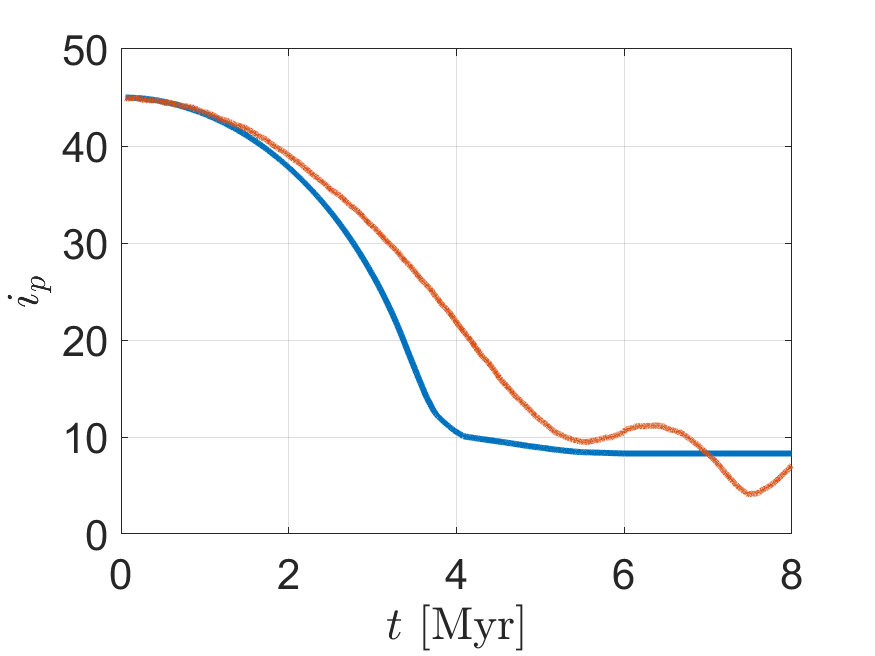}
  \caption{The predictions of equation \eqref{eqn:perturber's inclination equation of motion} in blue, compared with the numerical simulation (orange). The set-up and initial conditions are the same as figure \ref{fig:inclination histogram}.}
  \label{fig:early times}
\end{figure}

Next, we test the predictions of \S \ref{sec:early times} (and appendix \ref{sec: disc thickening}) for $i_{\rm p}$: we show, in figure \ref{fig:early times}, the solution of equation \eqref{eqn:perturber's inclination equation of motion}, for the same numerical set-up as in figure \ref{fig:inclination histogram}, and compare it with the simulation. While the two do not completely agree, they do so approximately, and one can see that the treatment of \S \ref{sec:early times} offers a way to understand the alignment of the inclination of the perturber with the disc.

\begin{figure*}
  \centering
  \includegraphics[width=\linewidth]{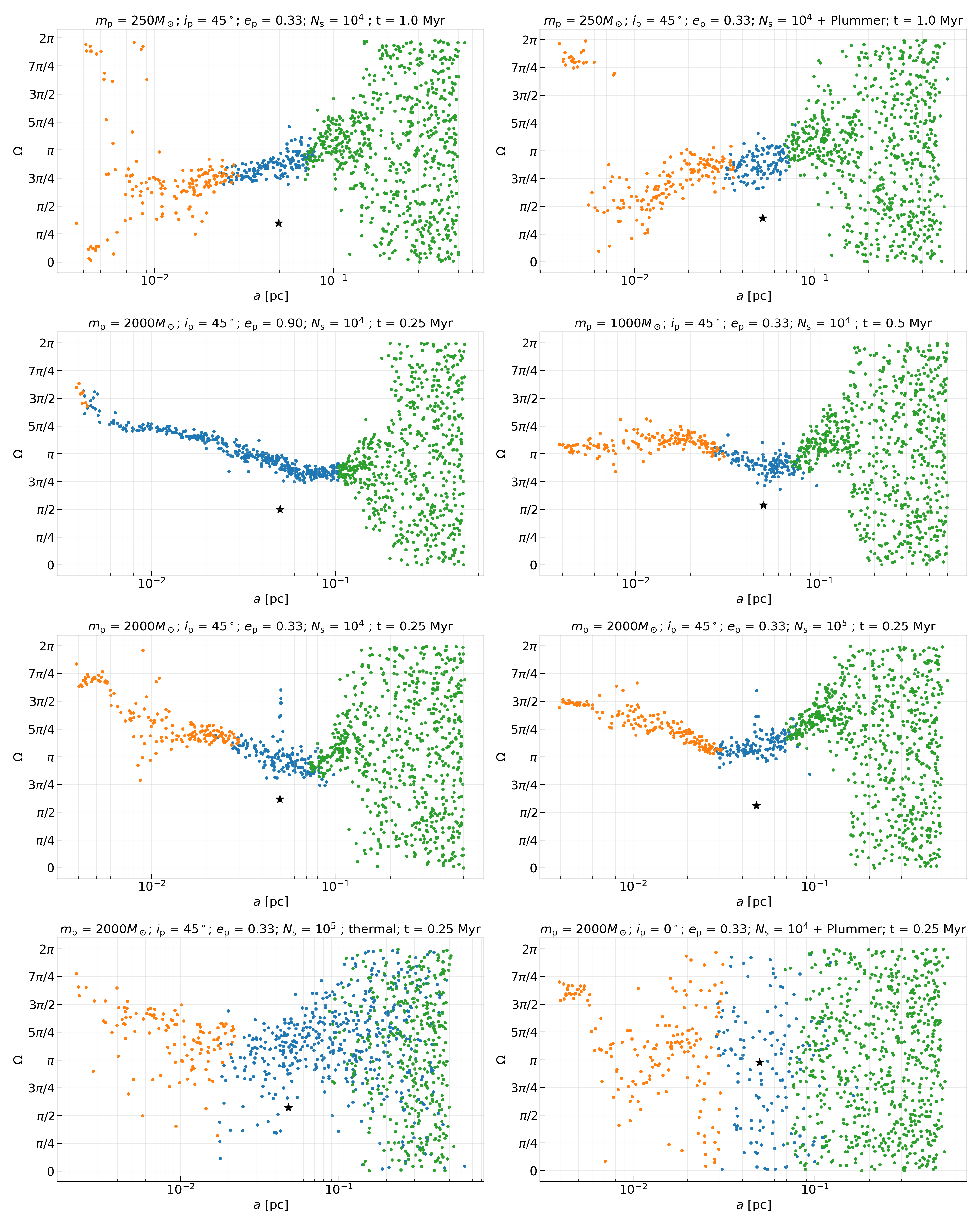}
  \caption{Same as Fig.~\ref{fig:omega histogram}, but for the models in the other rows of table \ref{tab:runs}. The title of each panel describes the initial parameters of the IMBH -- namely, its mass, inclination angle with respect to the initial disc plane, and its eccentricity, as well as the total number of particles in the simulation, and the time. `+ Plummer' in the title means that the stellar disc and a spherical component were embedded in an external Plummer potential. The word `thermal' in the title refers to the thermal eccentricity distribution for the stellar disc.}
  \label{fig:omega scatter all}
\end{figure*}

\begin{figure*}
  \centering
  \includegraphics[width=\linewidth]{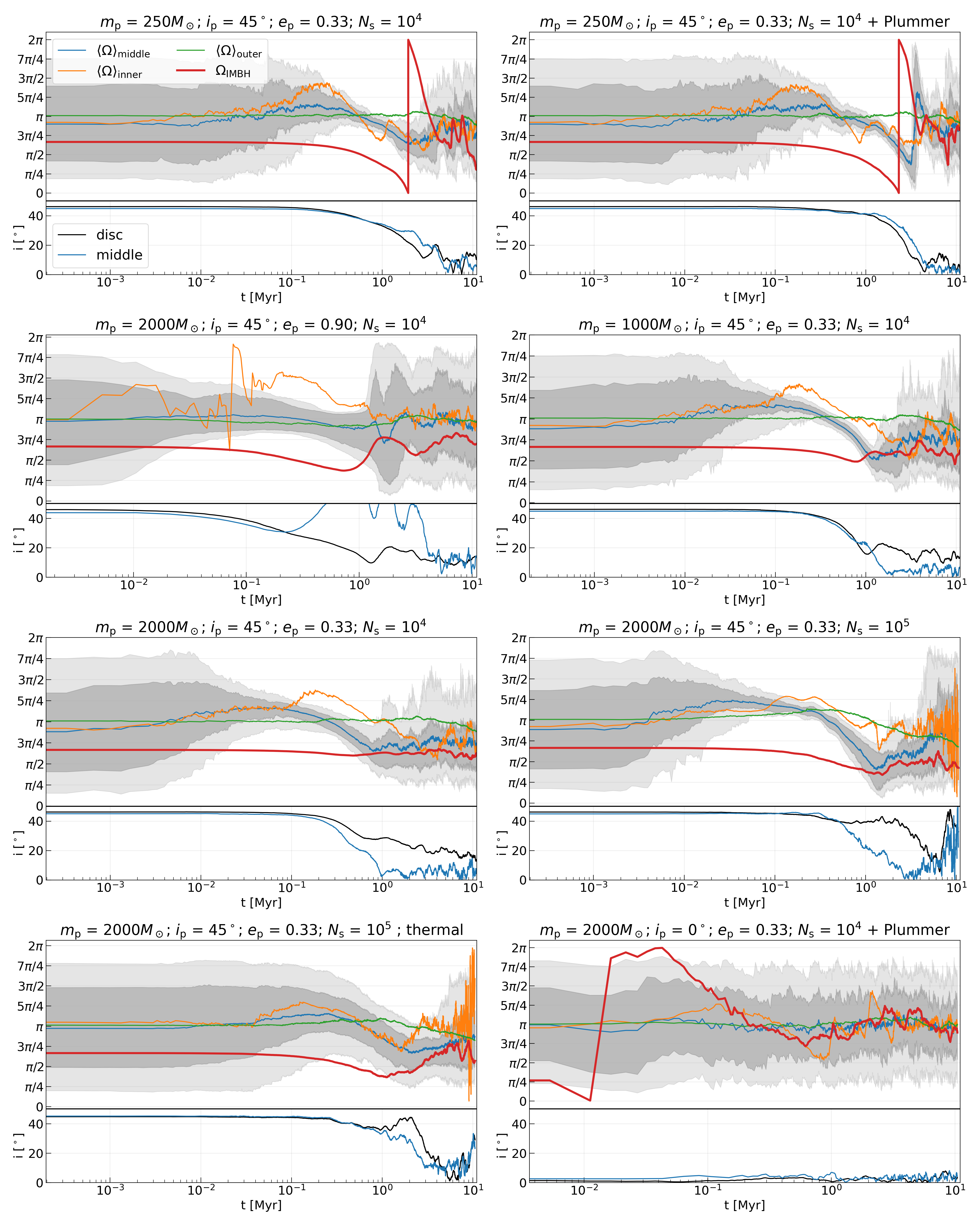}
  \caption{Same as Fig.~\ref{fig:omega evolution}, but for the models in the other rows of table \ref{tab:runs}. The order of the panels is the same as in figure \ref{fig:omega scatter all}.}
  \label{fig:omega t all}
\end{figure*}

We now proceed to show that the phenomena of \S \ref{sec:early times} are not peculiar to the specific initial conditions considered above, that is to those of the first row of table \ref{tab:runs}. We vary both the mass-ratio $\eps$ between the perturber and the disc, the latter's thickness and the treatment of the spherical part of the system; we also change the perturber's eccentricity and finally its initial inclination. The results are summarised in figures \ref{fig:omega scatter all} and \ref{fig:omega t all}.

The first alteration we consider is turning the spherical component of the system into a `live' sphere. The spherical component was a pure monopole, Plummer potential, above, and here, by sampling $N_s$ particles from a spherically symmetric power-law distribution, and allowing them to evolve and interact with all other particles, the `sphere' could have non-zero higher multipoles, and a non-zero net angular momentum. This is the case with all the simulations shown in figures \ref{fig:omega scatter all} and \ref{fig:omega t all}. From the first row of figure \ref{fig:omega scatter all} we see that in this case the node alignment between the disc and the perturber still persists, especially with the stars that have a similar semi-major axis to $a_{\rm p}$, until the perturber coalesces with the disc. The effect is weakened, because the `sphere' has a non-vanishing quadrupole moment by chance, which competes with the influence of the perturber. Upon increasing $m_{\rm p}$, the time it takes the perturber to reach the disc is decreased, but one can still see an alignment of the nodes before that. As the sphere becomes too massive (right panel of the penultimate row of the figures) the effect becomes somewhat less pronounced, and likewise for a thicker disc (bottom right panel).

Another noteworthy property is that, as one expects, if the perturber starts inside the disc (bottom right panel), there is no alignment, because then $\ham_{\rm p}$ is far from dominating the dynamics.

\begin{figure}
    \centering
    \includegraphics[width=0.45\textwidth]{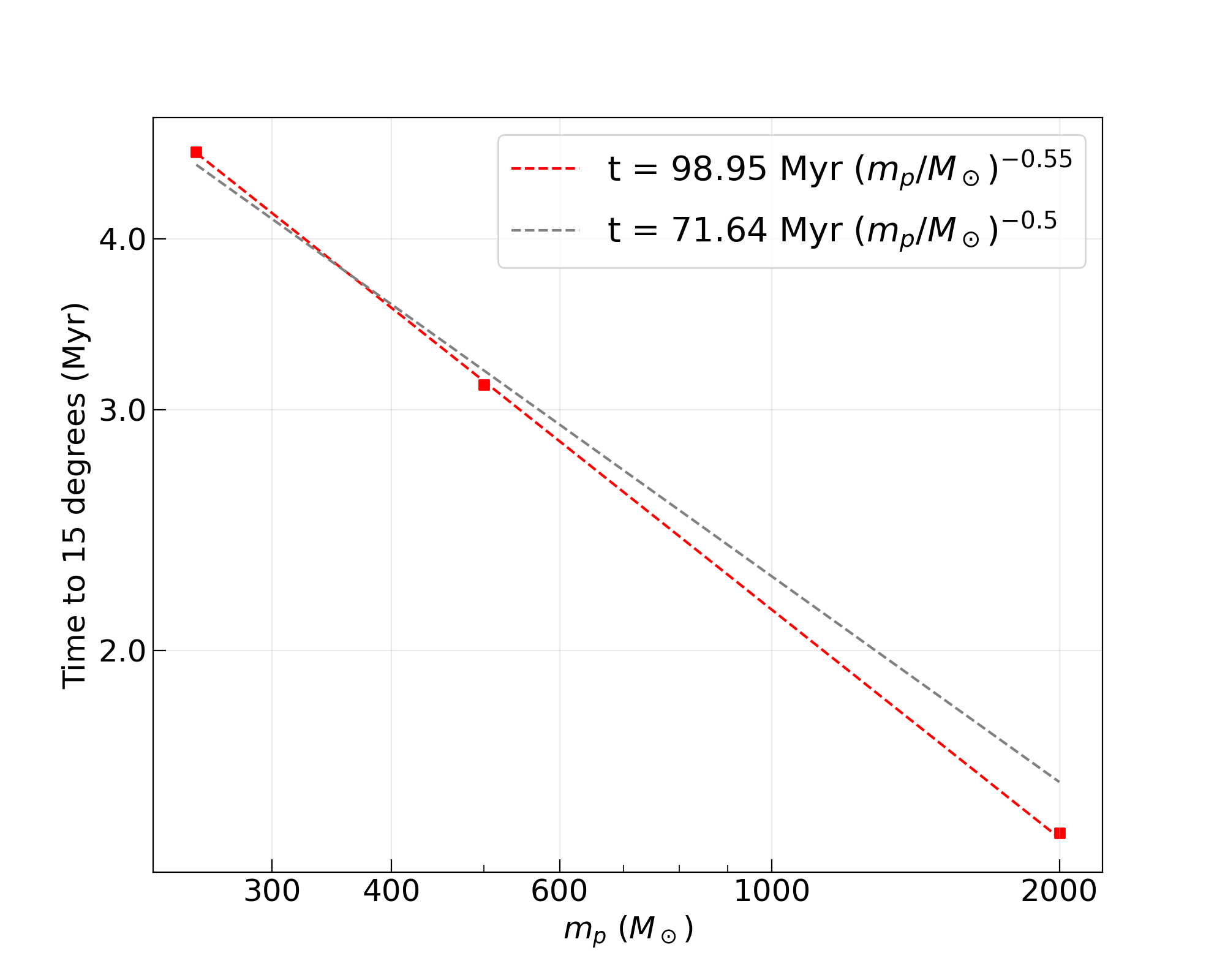}
    \caption{The total time it takes the perturber to decrease its inclination from $45^\circ$ to $15^\circ$ as a function of the perturber's mass. The dashed lines show the best fit (red) and the fit to the power-law of $-0.5$ (grey). The rest of the set-up is identical to that of figure \ref{fig:omega histogram}, as described in the text and in table \ref{tab:runs}.}
    \label{fig:alignment time scale}
\end{figure}

The alignment time-scale in equation \eqref{eqn:tau RDF} is proportional to $m_{\rm p}^{-1/2}$. In figure \ref{fig:alignment time scale} we test this dependence on the peturber's mass, by running the same simulation as in the first row of table \ref{tab:runs}, but with varying IMBH mass. The total time it takes the perturber -- starting at $i_{\rm p}(0) = 45^\circ$ -- to get to $i_{\rm p} = 15^\circ$ is seen to follow the theoretical prediction closely.

\section{Discussion}
\label{sec:discussion}

Having shown that the perturber induces a rapid alignment of the $\set{\Omega_n}$ with its argument of the ascending node, let us inquire what the most general conditions under which one can expect this phenomenon to arise are; that is, are the requirements that $m_{\rm p} \ll M_{\rm d}$, and that the disc stars' inclinations be much smaller than $\eps$ really necessary? Does the phenomenon persist when including higher multipoles? What if the perturber is not a single particle, but its contribution to $\ham$ is replaced by the shot noise fluctuations of a spherical stellar distribution, or a general $\ham_{\rm p}$?

In the most general setting, the full Hamiltonian may be decomposed as
\begin{equation}
    \ham = \ham_{\rm d} + \ham_{\rm p} + \ham_s,
\end{equation}
where $\ham_{\rm d}$ governs the self-interaction of the disc (which we assume is $\ham_{\rm LL}$), $\ham_s$ describes the interaction of the perturber with itself, and $\ham_{\rm p}$ includes all the interactions between the disc particles and the perturber(s). 
In the double orbit-averaged limit, the most general form for $\ham_{\rm p}$ we consider here is
\begin{equation}
    \ham_{\rm p} = \sum_{n,p,\ell} h_{pn\ell}P_{\ell}(\cos \theta_{pn}),
\end{equation}
where the index $n$ runs over the disc particles, $p$ refers to the perturber(s), and $\ell\in \mathbb{N}$ is the multipole index, starting at quadrupole, $\ell = 2$. The angle $\theta_{pn}$ is the angle between the angular momentum vector of perturbing particle $p$ and that of disc particle $n$. It is this angle which contains the entire dependence on $\Omega_n$, $i_n$, $\Omega_p$ and $i_p$. Let $\beta$ denote one of these four angles; we will restrict ourselves to the case where
\begin{equation}
    \frac{\partial P_{\ell}(\cos \theta_{pn})}{\partial \beta} \approx -\frac{\partial \theta_{pn}}{\partial \beta}\sin \theta_n P_{\ell}'(\cos \theta_n),
\end{equation}
where now $\theta_n$ is the angle between the angular momentum direction of particle $n$, and the total angular momentum of the perturbers $\hat{\mathbf{J}}_{\rm pert}$, i.e., we restrict ourselves to the case where the entire dependence on $p$ is in $\frac{\partial \theta_{pn}}{\partial \beta}$. This approximation is valid when the disc is thin. We may now write
\begin{equation}
    \hat{\mathbf{J}}_{\rm pert} = (\cos \Omega_{\rm pert} \sin i, \sin \Omega_{\rm pert} \sin i, \cos i),
\end{equation}
in complete analogy with the inclination and argument of the ascending node of the single perturber above.
By examining the way we derived equation \eqref{eqn: Omega_n alignment}, we see that it is necessary that $\dot{\Omega}_{\rm pert} \neq 0$, as well as that $\ham_{\rm p}$ dominate the dynamics of the disc particles. This is possible for the thin disc limit considered above.
If these two conditions hold, the derivation can be repeated, and a rapid alignment occurs.
For the single perturber case, as $\eps$ increases, both of these conditions are still met, but the disc thickens faster, over a time-scale $A_n$ (which is explicitly given in appendix \ref{sec: disc thickening}), and we know that the alignment persists only until the disc's thickness reaches $\ord{\eps}$.
Consequently, as the alignment occurs over a time-scale $b_{\textrm{p}n}^{-1}$, and the thickening requires time $A_n$, one must have
\begin{equation}
    A_n \ll b_{\textrm{p}n},
\end{equation}
as a necessary condition for $\Omega_n$ to align itself with $\Omega_{\rm p}$ in accordance with \eqref{eqn: Omega_n alignment}, i.e. precisely that the disc be thin, i.e. that inequality \eqref{eqn: alignment condition} be satisfied. These are the conditions under which such a phenomenon occurs, and it persists as long as they are met.\footnote{For $\eps \sim 1$, these conditions could be met, but the disc's thickness would grow too much over the course of its evolution, and hence the assumption of $\ham_{\rm d} = \ham_{\rm LL}$ would have to be modified at late times.}
Furthermore, it is also possible that for the outer extent of the disc, one would not have $\ham_{\rm p} \gg \ham_{\rm LL}$, and indeed, whether or not this alignment occurs varies from star to star.

Very recently \cite{Levin2022} studied a rotating spherical cluster and a (rotating) disc, which quickly align together around an SMBH; this alignment was derived from a statistical mechanical view-point using the fluctuation-dissipation theorem. If the perturber were replaced by a rotating cluster, then the discussion in the previous paragraph would apply. Indeed, the mass-scaling of the resonant friction time-scale derived by that work is the same as that in equation \eqref{eqn:tau RDF}.

This time-dependence is much faster that the ordinary $\tau_{\rm CDF} \sim M_\bullet^2/(m_{\rm p}M_{\rm d})\times t_{\rm orb}/\ln \Lambda$ one would find for Chandrasekhar dynamical friction. This implies that $\tau_{\rm RDF} \sim \sqrt{\tau_{\rm CDF} \, t_{\rm orb}\ln \Lambda}$ -- generally much shorter than $\tau_{\rm CDF}$. For our setting, $M_\bullet = 4\times 10^6 M_\odot$, $M_{\rm d} = 2000 M_\odot$, $m_{\rm p} \approx 250 M_{\odot}$, $t_{\rm orb} = 524$ years (for the same density-profile as in \S\ref{sec:simulation shortened} and in figure \ref{fig:time-scales}), we obtained
$\tau_{\rm RDF} \approx 4.9~\textrm{Myr}$, while $\tau_{\rm CDF} \approx 1.7~\textrm{Gyr}$ for $\ln \Lambda = 10$.
For smaller Coulomb logarithms, the latter becomes already of the order of the age of the Universe. For even more massive super-massive black holes, the difference is even more extreme: for instance for $m_{\rm p} = 1000 M_\odot$, $M_\bullet = 10^9 M_\odot$, and $M_{\rm d} = 10^4 M_\odot$, the orbital time is multiplied by $\sqrt{10^9/(4\times 10^6)}$, and we find $\tau_{\rm RDF} \approx 4.3~\textrm{Gyr}$ while $\tau_{\rm CDF} \approx 8.3\times 10^{13} ~\textrm{years} \gg H_0^{-1}$.
In other words, an alignment of the inclination due to resonant dynamical friction is possible in systems, where ordinary, Chandrasekhar, dynamical friction would take much too long to do so.

\subsection{Relevance to The Milky Way}
Given the estimated mass of the stellar disc in the Galactic centre of $M_\mathrm{d} \simeq 10^3-10^4 M_\odot$ \citep{Bartko2010}, a perturber of mass $m_{\rm p} \simeq 2.5\times10^2 - 10^4 M_\odot$ initially located above the disc plane ($i_{\rm p} \simeq 45^\circ$) may cause the alignment of the longitudes of the ascending nodes for the stars within the disc. Thus, the detection of a narrow spread in the longitudes of the ascending nodes within the stellar disc would be consistent with the presence of a perturber with semi-major axis similar to those of the stars that feature the alignment, while the mass of the potential perturber will depend on its position. An IMBH of such mass will decay into the disc within a few megayears by resonant dynamical friction, as one can see from figure \ref{fig:early times}.

\citet{Alietal2020} studied stars within the $S$-star cluster and young stellar disc with known orbital solutions and found anisotropies in the longitudes of the ascending node. This may be consistent with the existence of an IMBH with $a_{\rm p} \simeq 0.05$pc and mass less than the enclosed mass of the stellar disc in this region, but the features in $\Omega$ presented by \citet{Alietal2020} are more complex than the expectation from the effects of the dynamical friction of the massive perturber examined in this work. Scenarios with more than one IMBH, or the influence of higher multipoles, are, however, beyond the scope of this paper, and are the topic of future research. 

It is debated in the literature, whether a clumpy structure, known as the IRS 13 association \citep{Maillard2004} located at the projected distance of $\simeq0.13$pc from Sgr A$^*$ may host an IMBH of mass $\simeq1300 M_\odot$ \citep{PortegiesZwart2002}. Whether it harbours an IMBH or not, it will act as a perturber and, thus, may drive the alignment of the ascending nodes of the stars with overlapping orbits with IRS 13. Given the sharp transition in the distribution of $\Omega$ as a function of semi-major axis (presented in figure \ref{fig:omega scatter all}), if such transition is detected, one may use it to estimate the 3D distance to IRS 13 which is currently unknown \citep{Tsuboi2020}.

We should note that even in the absence of an IMBH, a live stellar halo gives rise to random quadrupole fluctuations, some of whose effects could be similar to those of an IMBH on a stellar disc. Indeed, a finite number of spherically distributed stars stochastically generates a shot-noise quadrupolar density fluctuation which leads to a torque proportional to $\sqrt{N\langle m^2 \rangle}$ which drives nodal precession for the disc stars \citep{Kocsis+2011}, identical to that of three IMBHs of masses of order $N\langle m^2 \rangle^{1/2}$ along the eigenvectors of the $V_{\alpha\beta}=\langle L_{\alpha}L_{\beta}\rangle$ tensor where $L_\alpha$ denotes the Cartesian components of angular momentum vectors of the spherically distributed stars \citep{KocsisTremaine2015,Roupas+2017}.\footnote{Here $N$ denotes the number of stars in the spherical distribution within the logarithmic radial bin of a test star, i.e. it stands for $\mathrm{d}N/\mathrm{d}\ln r$.}
These torques remain coherent over a time-scale of $t_{\rm coh} \propto t_{\rm orb} M_\bullet/\sqrt{N\langle m^2\rangle}$ \citep{KocsisTremaine2015,Fouvryetal2019}, which may be comparable to $\tau_{\rm RDF}$.\footnote{At later times, the eigenvectors reorient due to similar but misaligned quadrupole fluctations at other orbital radii and because of higher multipole fluctuations at the same orbital radii.} This suggests that shorter time-scale phenomena like the nodal alignment described in this paper could occur even without a massive perturber (see figure 20 of \citealt{Panamarev2022}). An in-depth exploration of this possibility is deferred to future work. In \citet{per+18} it was shown that a realistic live halo could create some structure in the disc, change its thickness, and even form spiral arms. Such over-densities might explain the origin of IRS 13 without the need for an IMBH.

It is also interesting to consider the possibility of several star-formation epochs, that give rise to multiple discs (see, e.g. \citealt{mas+19}) in which case the mutual interaction between the discs could resemble the effect of a massive perturber, although discs with comparable masses would not fulfil the requirements for the mass hierarchy discussed in the introduction, but they may apply to some extent if one of the discs is of significantly less massive than the other.

\subsection{Limitations}
In \S \ref{sec:early times} we solved the equations of motion approximately during the early times, when $\ham_{\rm p} \gg \ham_{\rm d}$, but eventually, when the disc thickens enough to be comparable to the perturber's inclination $\ham_s$ grows to be of a similar magnitude to $\ham_{\rm p}$. We accounted for that in the evolution of $s_n$ by including the limit $s_{n,{\rm late}}$. Afterwards, when $\ham_s$ dominates the dynamics, and $\ham_{\rm p}$ is a small perturbation, and if $\eps$ is small enough, the entire system $\ham_s + \ham_{\rm p}$ may be solved exactly in the Laplace-Lagrange approximation as in appendix \ref{subsec:oscillator}. Inclinations of more than $\pi/2$ are not explored in this paper, and are a topic of future work.

The equations of motion we solved here were derived from doubly-averaged Hamiltonians -- both over the mean anomalies and over the arguments of pericentre, so scalar resonant relaxation was not accounted for. We also did not include collisional effects like two-body relaxation or Chandrasekhar dynamical friction. These are all included in the numerical simulations, as they are direct $N$-body simulations. However there, the spherical component is treated as a potential in some cases, for computational reasons, and the number of particles is smaller than realistically. The spherical component is isotropic in angular momentum space, but the observational data indicate that the Milky Way nuclear star cluster has some net rotation \citep{Feldmeier2014}. Both the stellar disc and sphere in the simulations were treated as one stellar population of old stars while in reality the stellar disc consists of young stars with ages below 10 Myr \citep{LevinBeloborodov2003, PaumardEtAl2006, Yelda2014,Habibietal2017}. The stellar evolution effects are important for general understanding of the dynamics in the close vicinity of the SMBH, but we do not expect them to affect the alignment of longitudes of the ascending nodes presented here, as the alignment happens on much shorter time-scales. Nevertheless, a realistic mass function for the stellar disc, and stellar evolution aspects would play a role in the timescale for thickening the disc \citep{Mik+17}; disc thickening due to two-body interactions should take place on a time-scale similar to that of Chandrasekhar dynamical friction, which is much longer than $\tau_{\rm RDF}$ for $\eps \ll 1$. (For $\eps \approx 1$, these can start to become comparable: in the corresponding models the semi-major axis of an IMBH shrinks by the factor of two within 5 Myr.)

We have also neglected relativistic effects, i.e. the relativistic apsidal precession and Lense-Thirring precession. The latter changes $\Omega$ and $i$ very close to the central SMBH. Its contribution to the Hamiltonian is
$\ham_{\rm LT} = \sum_{n} h^{\rm LT}_n (\mathbf{S}_\bullet \cdot \mathbf{J}_n) + h^{\rm LT}_{\rm p} (\mathbf{S}_\bullet \cdot \mathbf{J}_{\rm p})$,
where $\mathbf{S}_\bullet$ is the angular momentum of the SMBH, and $h^{\rm LT}_{n, \textrm{p}}$ are constants. This will result in, e.g., contributions to the $\dot{\Omega}_n$ equations, which is expected to ruin the node alignment, if the Lense-Thirring time-scale, $\omega^{\rm LT}_n \equiv \frac{2GS_\bullet}{c^2a_n^3(1-e_n^2)^{3/2}}$, is smaller than $b_{{\rm p}n}(0)$. To account for the Lense-Thirring precession properly, one would have to include $\ham_{\rm LT}$ in the equations of motion, and modify the multi-scale asymptotic expansion to include this additional time-scale. As this time-scale would usually be longer than $\nu_{\rm p}$ and $\tau_{\rm RDF}$ (e.g. \citealt{KocsisTremaine2011}), this would generally manifest itself in the later stages of the evolution, i.e. after the perturber has entered the disc, i.e. when resonant dynamical friction is already weak. In other words, this relativistic effect would primarily act as a correction to the stages of the evolution where the Newtonian interactions of the entire system (disc + perturber) is already well-modelled by a Laplace-Lagrange Hamiltonian.

\subsection{Final state of relaxation}
In this paper we studied the alignment of the orbital plane of a point mass perturber with a massive disc from an initial highly misaligned configuration to the point where the perturber's inclination starts to overlap with that of the disc stars. The final RMS inclination angle may be obtained using statistical physics: in the limit in which  (i) a thin disc dominates the energy budget of the system, (ii) the net angular momentum is far from zero,\footnote{E.g. most stars orbit in the same direction.} and (iii)
the correlations between the angular momenta of stars are neglected other than the conservation of the total VRR energy and angular momentum,
then the RMS inclination angle of the perturber relative to those of disc stars is $\langle i_p^{2}\rangle^{1/2} = (m/m_p)^{1/2} \langle i^2\rangle^{1/2}$ \citep[see Eq. 35 in][]{Wang_Kocsis2023}. In this sense the perturber is ultimately expected to be confined very close to the mid-plane of the disc.

However, further study is required to verify the accuracy of this conclusion, as the no-correlation assumption (iii) might fail. In particular, \citet{Kocsis+2011} showed that in the thin disc limit, the angular momentum vectors exhibit long-range spatial correlations. In particular, the disc behaves as a system of harmonic oscillators, which undergo normal mode oscillations. The angular momentum vector directions oscillate with two degrees of freedom about the mean angular momentum of the system. The normal mode oscillation amplitudes are independent, but the inclination angles of individual stars are correlated as they are components of the modes.
The normal mode oscillation amplitudes are set by requiring that each normal mode is at the same temperature $T$ and rotation temperature. Here `temperature' is the inverse Lagrange multiplier that enforces the conservation of total energy when maximising entropy, as usual, and rotation temperature is the analogous Lagrange multiplier  that enforces the conservation of total angular momentum. The energies of the normal modes generally do not obey equipartition,\footnote{Equipartition holds only if the net angular momentum is zero, in that case each mode has energy $k_{\rm B}T$.} but they may either grow or decrease with the mode oscillation frequency, depending on the ratio of total energy to angular momentum. A further complication is that the perturber is coupled to the stellar cluster through more than one normal modes. We leave a detailed study of the possible range of final RMS inclination angles of the perturber as a function of the disc properties to future work.

\section{Summary and Conclusions}
\label{sec:summary}
In this paper we developed an analytical model, based on resonant relaxation and singular perturbation theory, which captures the essential features of resonant dynamical friction. We showed that the singular nature of the equations of motion implies, in the case of a thin disc, a rapid alignment of the arguments of the ascending node of the perturber and the disc particles, which then gives rise to a coherent torque, which aligns the perturber with the disc. We showed that the predictions of this model mesh well with results of $N$-body simulations, and found that the node alignment occurs for a wide range of initial conditions. In particular, one may safely conclude that the perturber would not re-orient to the disc if the disc particles were treated as test-particles, for it is their gravity that sources an $\ord{1}$ value of $\dot{\Omega}_{\rm p}$, but on the other hand, the perturber's contribution to the potential does dominate their motion, at least initially.

The instantaneous alignment timescale $(\mathrm{d}i_{\rm p}/\mathrm{d}t)^{-1}$ is proportional to $m_{\rm p}^{-1}$, but the total alignment time-scale as a function of SMBH, IMBH, and local disc mass $(M_{\bullet},m_{\rm p},M_{\rm d,loc})$ is proportional to $M_{\bullet}/(m_{\rm p}M_{\rm d,loc})^{1/2}$ -- much faster than Chandrasekhar dynamical friction timescale which scales as $M_{\bullet}^2/(m_{\rm p}M_{\rm d,loc})$.
We expect some of the results of this analysis to extend to more general perturbers of the disc, such as a `live' spherical component instead of a single point-particle perturber, or the case of two stellar discs, or to certain planetary systems.

\section*{Acknowledgements}
We wish to thank the anonymous referee for reviewing our paper insightfully. We are grateful to Yuri Levin and Mor Rozner for helpful comments on the manuscript.
Y. B. G. is grateful for the kind hospitality of the Rudolf Peierls Centre for Theoretical Physics at the University of Oxford and to St. Hugh's College, Oxford, where some work on this research was done. Y. B. G. is supported by the Adams Fellowship Programme of the Israeli Academy of Sciences and Humanities.
T. P. and B. K. received funding from the European Research Council (ERC) under the European Union’s Horizon 2020 Programme for Research and Innovation ERC-2014-STG under grant agreement No. 638435 (GalNUC), and were also supported by the Science and Technology Facilities Council (STFC) Grant Number ST/W000903/1. T. P. acknowledges the support of the Science Committee of the Ministry of Science and Higher Education of the Republic of Kazakhstan (Grant No. AP14870501).

\section*{Data Availability}
The data underlying this article will be shared on reasonable request to the corresponding author.




\bibliographystyle{mnras}
\bibliography{friction_bib}



\appendix

\section{Solution of The Equations of Motion}
\label{appendix: equations of motion}

The purpose of this appendix is to provide a detailed solution of the equations of motion, justifying the statements made in \S \ref{sec:early times}. As explained in \S \ref{sec:set-up}, the relevant Hamiltonian is $\ham = \ham_{\rm p} + \ham_{\rm LL}$, where $\ham_{\rm p}$ is defined in equation \eqref{eqn:ham p definition}; $\ham_{\rm LL}$ governs the disc's self-interactions. One can define a pair of oblique canonical variables as in, e.g., \citet{KocsisTremaine2011}
\begin{align}\label{eqn:definition p_n q_n}
    q_n \equiv \gamma_n s_n \sin \Omega_n, & & p_n \equiv -\gamma_n s_n \cos\Omega_n,
\end{align}
where $s_n \equiv \sin i_n$, . Then, $\ham_{\rm LL}$ is actually the Hamiltonian of a collection of coupled harmonic oscillators:
\begin{equation}\label{eqn: ham LL definition}
  \ham_{\rm LL} = p_n A_{nm} p_m + q_nA_{nm}q_m,
\end{equation}
with the constant matrix $A_{nm}$ defined by equation (40) of \citet{KocsisTremaine2011}, \emph{viz.}
\begin{equation}\label{eqn: Anm definition}
  A_{nm} = \begin{cases}
             -\frac{Gm_nm_m\alpha_{nm}b_{3/2}^{(1)}(\alpha_{nm})}{8\max{\set{a_n,a_m}}\gamma_n\gamma_m}, & \mbox{if } n\neq m \\
             \sum_{k\neq n}\frac{Gm_nm_m\alpha_{nm}b_{3/2}^{(1)}(\alpha_{nm})}{8\max{\set{a_n,a_m}}\gamma_n\gamma_m}, & \mbox{otherwise}.
           \end{cases}
\end{equation}

The full equations of motion may be obtained from $\ham$ in the usual way:
\begin{align}
  \label{eqn:i n dot Lagrange equation of motion}\frac{\mathrm{d}i_n}{\mathrm{d}t} & = -\frac{1}{\gamma_n^2\sin i_{n}}\frac{\partial \ham_{\rm p}}{\partial \Omega_n} -\frac{1}{\gamma_n^2\sin i_{n}}\frac{\partial \ham_{\rm LL}}{\partial \Omega_n} \\ &
  = - \frac{Gm_{\rm p}m_n}{\gamma_n^2\max\set{a_{\rm p},a_n}} \sin i_{\rm p} \sin(\Omega_{\rm p} - \Omega_n) \\ &
  \times \sum_{l=2}^\infty P_{\ell}(0)^2 s_{pnl}\alpha_{pn}^l P_{\ell}'(\cos\theta_{\textrm{p}n}) -\frac{1}{\gamma_n^2\sin i_{n}}\frac{\partial \ham_{\rm LL}}{\partial \Omega_{n}} \\
  \frac{\mathrm{d}\Omega_n}{\mathrm{d}t} & = -\frac{1}{\gamma_n^2\sin i_n}\frac{\partial \ham_{\rm p}}{\partial i_n} - \frac{1}{\gamma_n^2\sin i_n}\frac{\partial \ham_{\rm LL}}{\partial i_n} \\ &
  = \frac{Gm_{\rm p}m_n}{\gamma_n^2\max\set{a_{\rm p},a_n}} \left[\sin i_{\rm p}\cot i_n \cos(\Omega_{\rm p} - \Omega_n) -\cos i_{\rm p}\right] \\ &
  \times \sum_{l=2}^\infty P_{\ell}(0)^2 s_{pnl}\alpha_{pn}^l P_{\ell}'(\cos\theta_{\textrm{p}n}) - \frac{1}{\gamma_n^2\sin i_n}\frac{\partial \ham_{\rm LL}}{\partial i_n} ,\label{eqn:omega n dot Lagrange equation of motion}
\end{align}
where
\begin{align}
    \gamma_n &\equiv \sqrt{\mu_n\sqrt{G(m_n + M_\bullet)a_n(1-e_n^2)}} \nonumber\\
    &\approx \sqrt{m_n\sqrt{G M_\bullet a_n(1-e_n^2)}}
    \,,
\end{align}
with the reduced mass defined as $\mu_n \equiv m_nM_\bullet/(M_\bullet + m_n)\approx m_n$. Let us solve the equations of motion perturbatively.

\subsection{Zeroth-Order Solution}
\label{subsec:zeroth order}

Suppose one starts with initial conditions where $i_n = 0$ at $t=0$, and $\Omega_n$ is `randomly' distributed between $0$ and $2\pi$. Then, since $\ham_{\rm LL}$ is proportional to $i_n i_m$, to leading order in $\set{i_n}$, both $\Omega_n$ and $i_n$ are conserved for these initial conditions. This also implies that so are $i_{\rm p}$ and $\Omega_{\rm p}$ to zeroth order in $\eps$.

To find any non-trivial behaviour one has to go to higher order in $\eps$. Here, despite approximating the disc's self-interaction by a Laplace-Lagrange Hamiltonian, we allow $i_{\rm p}$ to be arbitrarily large; for $\ham_{\rm p}$, we truncate the multipole expansion at the quadrupole. Let us start by trying to solve the equations of motion iteratively. The quadrupole piece in $\ham_{\rm p}$ is
\begin{equation}
  \ham_{\rm p} \approx -G\sum_{n=1}^{N}\frac{m_{\rm p}m_n s_{pn2}}{8\max\set{a_{\rm p},a_n}}(3\cos^2\theta_{pn} - 1)
\end{equation}
where $\cos\theta_{pn}$ is given by Eq.~\eqref{eq:cos(theta_pn)}.
Now, since the zeroth order solution is $i_n = 0$, if one simply inserts that into $\ham_{\rm p}$, then $\cos\theta_{pn} = \cos i_{\rm p}$, whence $i_{\rm p}$ is a constant of motion, and $\Omega_{\rm p}$ now evolves linearly in time,
\begin{equation}\label{eqn:omega p zeroth order}
  \Omega_{\rm p}(t) = \nu_{\rm p} t + \Omega_{\rm p0},
\end{equation}
with $\ord{1}$ frequency
\begin{equation}
    \nu_p = -\frac{3}{4}n_{\rm p}\cos(i_{\rm p })\sum_{n=1}^N \frac{m_n}{M_\bullet}\frac{a_{\rm p}s_{pn2}\alpha_{pn}^2}{\max\set{a_p,a_n}},
\end{equation}
and initial condition $\Omega_{\rm p0} = \Omega_{\rm p}(t=0)$. The singularity of the equations manifests itself in that $0 \neq \dot{\Omega}_{\rm p} = \ord{1}$ for any $\eps >0$ (because $\eps$ in the numerator cancels with $\gamma_{\rm p}^2$ in the denominator), but $\dot{\Omega}_{\rm p} = 0$ for $\eps = 0$. Recall that at this order, $i_{\rm p}$ is constant, and equal to its initial value.

For example, for a power-law surface density profile $\Sigma(r) = \frac{M_{\rm d}(3-\beta)}{2\pi R^2}(R/r)^\beta$, for $0 \leq r \leq R$, $\beta < 2$, and where $a_{\rm p} \leq R$, we find, for circular orbits,
\begin{equation}
    \nu_{\rm p} = -\frac{3(2-\beta)}{4} \frac{M_{\rm d}}{M_\bullet}\frac{a_{\rm p}^2}{R^2}\left[\left(\frac{R}{a_{\rm p}}\right)^\beta + \frac{1-a_{\rm p}/R}{1+\beta}\right] \times n_{\rm p}.
\end{equation}

\subsection{Interlude -- Boundary Layers}
\label{sec:boundary layer}
\begin{figure}
    \centering
    \includegraphics[width=0.48\textwidth]{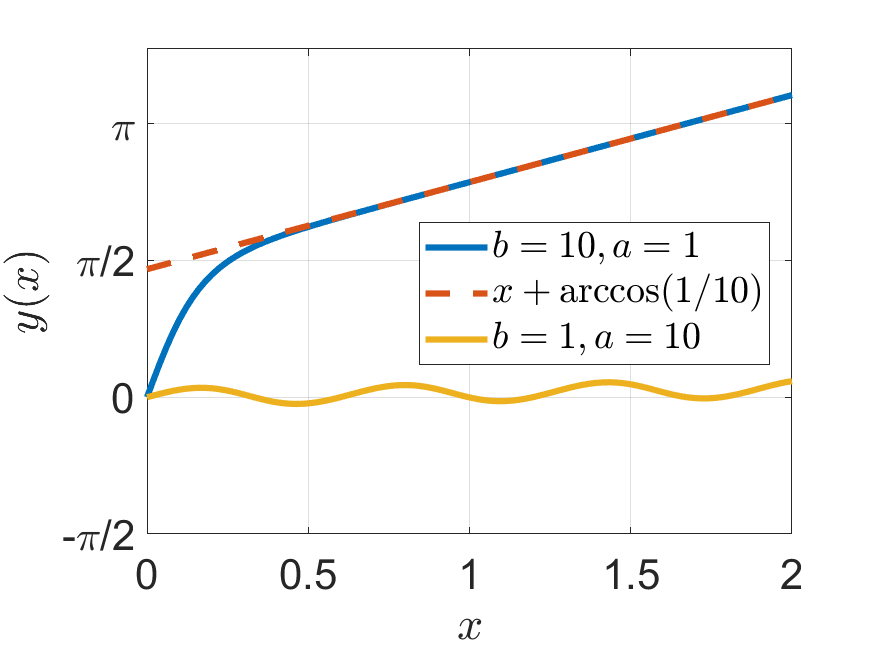}
    \caption{The solution of equation \eqref{eqn:example ode} for the first case, with $a=1$, $b=10$ (in blue), as well as its asymptotic limit $x+\arccos(1/10)$ (orange, dashed), and the other case, with $a=10$, $b=1$ (in yellow). The initial condition is $y(0) = 0$.}
    \label{fig:example ode solution plot}
\end{figure}

Before proceeding, let us study a related ordinary differential equation which displays the same type of singularity as the equations of motion (\ref{eqn: omega n dot main text} or \ref{eqn:Hamilton equation Omega n} below), but can be solved analytically:
\begin{equation}\label{eqn:example ode}
    \frac{\mathrm{d}y}{\mathrm{d}x} = b \cos(a x - y).
\end{equation}
This equation can be solved by substituting $u(x) = a x - y(x)$;
there are two options for the asymptotic behaviour as $x\to \infty$:
\begin{equation}\label{eqn:example ode limits}
    y \underset{x\to \infty}{\to} \begin{cases}
    ax + \arccos\left(\frac{a}{b}\right), & \mbox{if } \abs{a} \leq \abs{b} \\
    \mbox{oscillations}, & \mbox{otherwise}
    \end{cases}.
\end{equation}
One can understand this behaviour qualitatively as follows: if $\abs{a} > \abs{b}$, then $y$ cannot align itself with $ax$, simply because the cosine is too small. Indeed, if $b \gg a$, the cosine oscillates so fast that $y'$ changes sign all the time, and $y$ just fluctuates about its initial value. On the other hand, if $\abs{a} \leq \abs{b}$, the cosine is sufficiently large to allow $y$ to align itself with $ax$, and that is indeed what happens. A plot of the exact solution to equation \eqref{eqn:example ode} in the two cases is given in figure \ref{fig:example ode solution plot}.

For the first case, $y$ settles on its asymptotic value after an interval of order $\sim \abs{\frac{a}{b}}$ about $0$. Therefore, if $a \ll b$, this happens very quickly, i.e. \emph{there is a boundary layer at} $x = 0$ of thickness $\sim \frac{a}{b}$.
Generally, a boundary layer is a thin layer, over which a solution to a differential equation involving some small parameter $\delta$ changes very rapidly (e.g. a layer of size $\ord{\delta}$ over which the solution changes by an $\ord{1}$ amount), because of the incompatibility of the boundary conditions with the $\delta = 0$ limit, because the order of the equation decreases when $\delta = 0$ (see, e.g., \citealt{Verhulst2005} for a precise definition). In the above example, for instance, if $b = a/\delta$, setting $\delta = 0$ would imply that a generic initial condition $y(x = 0) = y_0$ is in general incompatible with the boundary condition $y = \arccos(a/b)$ at $x = \infty$. Below we denote the solution inside/outside the boundary layer as the `inner'/`outer' solutions, respectively.

Consider now the more general case of
\begin{equation}
    \frac{\mathrm{d}y}{\mathrm{d}x} = \frac{1}{\delta}\sum_{k=1}^K b_k \cos[g_k(x) - y] + f(x,y),
\end{equation}
where $\set{b_k}$, $\set{g_k}$ and $f$ are $\ord{1}$, and $\delta \ll 1$, with the initial condition $y(0) = y_0$. Again, it can be shown that there is a boundary layer at $x=0$ of thickness $\delta$.\footnote{Note that for a generic initial condition, the boundary layer will occur wherever we set the initial condition.} Let us study the case where $K=1$ in more detail:
we have
\begin{equation}
    \frac{\mathrm{d}y}{\mathrm{d}x} = \frac{b}{\delta} \cos[g(x) - y] + f(x,y).
\end{equation}

For $x\sim \ord{\delta}$, the derivative is also of order $\delta^{-1}$, so we use the multiple-scale method (for a general reference see, for instance, \citealt{PavliotisStuart2008}, and, e.g. \citealt{Ginat2021,Will2021} for applications in astrophysics), by defining $X \equiv x/\delta$, and treating $X$ and $x$ as independent variables, i.e. by replacing
\begin{equation}
    \frac{\mathrm{d}}{\mathrm{d}x} \mapsto \frac{\partial}{\partial x} + \frac{1}{\delta}\frac{\partial}{\partial X}.
\end{equation}
The equation then becomes
\begin{equation}
    \frac{\partial y}{\partial x} + \frac{1}{\delta}\frac{\partial y}{\partial X} = \frac{b}{\delta} \cos[g(x) - y(x,X;\delta)] + f(x,y).
\end{equation}
Let us expand
\begin{equation}
    y(x,X;\delta) \sim y_{(0)}(x,X) + \delta y_{(1)}(x,X) + \mbox{h.o.t.},
\end{equation}
where $\mbox{h.o.t.}$ denotes higher order terms,
and solve for $y_{(0)}$. The leading-order equation ($\ord{\delta^{-1}})$ is
\begin{equation}
    \frac{\partial y_{(0)}}{\partial X} = b\cos[g(x) - y_{(0)}].
\end{equation}
This is solved by the leading-order inner solution 
\begin{equation}
    y_{(0)}(x,X) = 2\arctan\left(A(x) e^{bX} \right) + g(x) - \frac{\pi}{2},
\end{equation}
where $A$ is an unknown function (determined by requiring the solvability of the next-order equation). For us, all that matters is that one can set $A(0)$ by requiring
\begin{equation}
    y_0 = 2\arctan\left[A(0)\right] + g(0) - \frac{\pi}{2}.
\end{equation}
In the limit $X \to \infty$, we find that
\begin{equation}
    y_{(0)} \to g(x) + \frac{\pi}{2}.
\end{equation}
Thus, in the limit $\delta \ll 1$, we found a leading-order uniform solution  $y_{(0)}(x,x/\delta)$, which approaches $g(x)+\pi/2$, almost immediately, after an $\ord{\delta}$ interval beyond $x=0$, given by
\begin{equation}
    y_{(0)}(x) = 2\arctan\left(A(x) e^{bx/\delta} \right) + g(x) - \frac{\pi}{2}.
\end{equation}

\subsection{Rapid Alignment of Nodes}
\label{sec:alignment}
Let us now return to the physical question, and show that if the initial inclinations of the disc stars are such that $\sin i_{n}(0) \ll \eps$, then the equations of motion for $\Omega_n$ exhibit a boundary layer behaviour, where $\Omega_n$ rapidly aligns itself with $\Omega_{\rm p}$, up to a phase, such that
\begin{equation}
  \cos (\Omega_{\rm p}(t) - \Omega_n(t)) = \frac{-\nu_{\rm p}}{b_{{\rm p}n}},
\end{equation}
where we define
\begin{align}
  b_{{\rm p}n} & \equiv  A_{n}\frac{\cos^2 i_n}{\sin i_{n}}, \\
A_{n} & \equiv  \frac{3}{8}\frac{Gm_{\rm p}m_ns_{pn2}\alpha_{pn}^2}{\gamma_n^2 \max\set{a_{\rm p},a_n}} \sin2i_{p} \\ &
= \frac{3}{8}n_n\,\frac{m_{\rm p}}{M_\bullet} \frac{a_n}{\max\set{a_{\rm p},a_n}} s_{pn2}\alpha_{pn}^2\,\sin2i_{p}.
\end{align}
This rapid alignment occurs because of the equation of motion for $\Omega_n$ is
\begin{equation}\label{eqn:Hamilton equation Omega n}
  \dot{\Omega}_n = \frac{1}{\gamma_n^2 }\frac{\partial \ham_{\rm p}}{\partial (\cos i_n)} = b_{{\rm p}n}\cos\left(\Omega_{\rm p} - \Omega_n\right),
\end{equation}
where terms from $\ham_{\rm LL}$ are sub-leading as long as $\sin i_n \ll \eps$. At this order, we also set $\cos \theta_{\textrm{p}n} \approx \cos i_{\rm p} \cos i_n$ inside the $P_{\ell}'$ in equation \eqref{eqn:omega n dot Lagrange equation of motion}, and neglected $\cos i_{\rm p}$ relative to $\sin i_{\rm p} \cot i_n \cos (\Omega_n - \Omega_{\rm p})$, because $\abs{\cot i_n} \gg 1$.
This equation has precisely the form discussed in \S \ref{sec:boundary layer}; explicitly, the mapping is $(a,-b,x,y) \mapsto (\nu_{\rm p},b_{{\rm p}n},t,\Omega_n)$. For $b_{\textrm{p}n} \geq \nu_{\rm p}$, the solution $\Omega_n$ changes, after a time $\propto \nu_{\rm p}/b_{\textrm{p}n} \leq 1$
-- a boundary layer -- such that the argument of the cosine remains constant, \emph{viz.}
\begin{equation}
  \cos(\Omega_n - \Omega_{\rm p}) = \frac{-\nu_{\rm p}}{b_{\textrm{p}n}}.
\end{equation}
This is exactly the type of phenomenon discussed above, and it implies that any perturbative treatment of this problem is deemed to fail, unless one accounts for the boundary layer properly. The alignment time-scale is $\sim \nu_{\rm p}^{-1} \times (\nu_{\rm p}/b_{\textrm{p}n}(0)) = 1/b_{\textrm{p}n}(0)$. Importantly, it is proportional to $m_{\rm p}^{-1}$.

Solving equation \eqref{eqn: Omega_n alignment} yields that
\begin{equation}
    \Omega_n(t) = \Omega_{\rm p}(t) + \arccos\left(\frac{-\nu_{\rm p}}{b_{\textrm{p}n}}\right) \approx \Omega_{\rm p} + \frac{\pi}{2} + \frac{\nu_{\rm p}}{b_{\textrm{p}n}},
\end{equation}
which is exactly equation the alignment described in the main text.

The above solution is correct to leading order. For completeness we note that the analysis may be extended to next-to-leading order as follows\footnote{we restrict attention to the leading order solution elsewhere in this paper}: we may write $\Omega_n(t) = \Omega_n^{(0)} + \frac{\nu_{\rm p}}{b_{{\rm p}n}}\Omega_{n}^{(1)} + \ldots$. Then, substituting this into equation \eqref{eqn:omega n dot Lagrange equation of motion}, one finds
\begin{align}
  \Omega_n & = \Omega_n^{(0)} + \frac{\nu_{\rm p}}{b_{{\rm p}n}}\Omega_{n}^{(1)} = \Omega_{\rm p} + \frac{\pi}{2} + \frac{\nu_{\rm p}}{b_{{\rm p}n}} - \frac{\cot i_{\rm p}}{\cot i_n} \\ &
   - \frac{\gamma_n^{-2}\nu_{\rm p}}{A_n\cos^2i_n}\left[\frac{\partial \ham_{\rm LL}}{\partial i_n}\right]_{\Omega_m = \Omega_{\rm p} + \frac{\pi}{2}, \;\forall m\in\set{1,\ldots,N}} + O\left[\frac{\nu_{\rm p}^2}{b_{{\rm p}n}^2}\right],
\end{align}
Or, upon substituting $\ham_{\rm LL}$,
\begin{equation}
  \begin{aligned}
    \Omega_n & = \Omega_{\rm p} + \frac{\pi}{2} + \frac{\nu_{\rm p}}{b_{{\rm p}n}} - \frac{\cot i_{\rm p}}{\cot i_n} \\ &
    - \frac{Gm_n\nu_{\rm p}\sin i_n}{8\gamma_n^{2}A_n\cos^2i_n}\sum_{k\neq n}\frac{m_k\alpha_{kn}b_{3/2}^{(1)}(\alpha_{kn})}{\max\set{a_k,a_n}} + O\left(\frac{\nu_{\rm p}^2}{b_{{\rm p}n}^2}\right).
  \end{aligned}
\end{equation}

\section{Harmonic Oscillator}
\label{subsec:oscillator}
To gain some more intuition, let us see what happens when both the disc's initial thickness is small, and the perturber's initial inclination is also small, such that additionally $i_n \ll i_p \ll 1$, and the Laplace-Lagrange approximation \citep[\S 7.7]{murray_dermott_2000} applies to the entire Hamiltonian.

In this appendix, we simply amalgamate $\ham_{\rm p}$ into $\ham_{\rm LL}$ by letting $n$ run from $1$ to $N+1$, where the $N+1$-st particle is the perturber, denoted by the $p$ index below. Note that in equations \eqref{eqn: ham LL definition} and \eqref{eqn: Anm definition}, $A_{np}\propto m_n^{1/2} m_p^{1/2}$, $A_{pp}\propto M_{\rm d}$ is a linear combination of all $m_n$ excluding $m_p$, and similarly  $A_{nn}$ is a linear combination excluding $m_n$ but including $m_p$.
As in \citet{KocsisTremaine2011}, one has to assume that the eccentricities and inclinations satisfy
\begin{equation}\label{eqn:condition Kocsis Tremaine 2011}
    e_n, s_n \ll \frac{\Delta a_n}{a_n}
\end{equation}
for $n \in \set{1,\ldots,N,p}$, in order to approximate the Hamiltonian as a harmonic oscillator, where $\Delta a_n$ is the difference between the semi-major axes of adjacent stars. Hamilton's equations of this Hamiltonian may be solved exactly by diagonalising the positive semi-definite matrix $A_{nm}$, and then decomposing the motion in its normal modes. But it is instructive to proceed somewhat differently, as we will do now.

The Laplace coefficient $b^{(1)}_{3/2}(\alpha_{nm})$ appears in the definition of $A_{nm}$, and therefore, if the nearly circular disc stars are too closely packed, then they will mask the effect of the perturber -- as $b^{(1)}_{3/2}(\alpha) \to \infty$ as $\alpha \to 1$ -- especially if stars are too close to the perturber as well. (See, e.g. \citealt{murray_dermott_2000} for explicit expressions for the Laplace coefficients.) For infinitesimally small $s_n$, this does not pose a problem, but in order for the approximation below we will need to require $b^{(1)}_{3/2} s_n \ll s_p$.\footnote{In calculating the Laplace coefficients we do not use any softening (cf. \citealt{SefilianRafikov2019}), because we have a finite number of particles, with no exactly overlapping orbits -- this is a good approximation to the dynamics when the disc thickness is sufficiently small such that $b^{(1)}_{3/2} s_n \ll s_p$.}

Introducing the complex phase space $z_n \equiv q_n + \mathrm{i} p_n = \gamma_n s_n e^{\mathrm{i}(\Omega_n -\frac{\pi}{2})}$, the equations of motion are
\begin{equation}\label{eqn:equations of motion perturber Harmonic oscillator}
    \dot{z}_n = -2\mathrm{i} \sum_{m=1}^{N+1} A_{nm} z_m,
\end{equation}
or equivalently
\begin{align}
    \dot{\Omega}_n &= -2 \sum_{m=1}^{N+1} A_{nm}\frac{\gamma_p s_p}{\gamma_n s_n}\cos(\Omega_m - \Omega_n)\nonumber\\
    &\approx -2 A_{np}\frac{\gamma_p s_p}{\gamma_n s_n}\cos(\Omega_p - \Omega_n) \label{eqn:omega n dot oscillator approximation},
\end{align}
and
\begin{align}
    \dot{s}_n &=-2 \sum_{m=1}^{N+1} A_{nm}\frac{\gamma_m s_m}{\gamma_n} \sin(\Omega_m - \Omega_n)\nonumber\\
    &\approx 2 A_{np}\frac{\gamma_p s_p}{\gamma_n} \sin( \Omega_n-\Omega_p),
\end{align}
where the approximations are valid for the stars ($1\leq n\leq N$) as long as $s_n \ll (m_n m_p)^{1/2} M_{\rm d}^{-1} s_{p}$ and $s_n\ll \Delta a_n/a_n$, as in this case the $A_{np}$ terms dominate the sum.

Equation \eqref{eqn:omega n dot oscillator approximation} is the same as equation \eqref{eqn:Hamilton equation Omega n}, as $A_{np} < 0$. Hence, we expect that the same alignment occurs, as long as the coefficient of the cosine remains large.
Observe that while, in this approximation, $s_p$ is time-independent, $s_n$ is not necessarily so. Indeed, direct computation from the above equations yields $p_n^2 + q_n^2 \approx C^2A_{np}^2/A_{pp}^2 + 2 A_{np}C\left[r_n\cos(\phi-\nu_{p}t)+k_n\sin(\phi-\nu_{\rm p}t)\right]/A_{pp}$, where $C$, $r_n$, $k_n$ and $\phi$ are constants.
which implies that even if $s_n$ is arbitrarily small ($\ll s_p m_{\rm p}/M_{\rm d}$ at time $t=0$, whence the coefficient of the above equation is very large), it rises to a magnitude $\frac{m_p}{M_{\rm d}}s_p$ after a time $\sim \abs{\nu_{p}}^{-1}$, whereupon the alignment stops (as the coefficient of the cosine becomes unity), and the interaction with the other disc stars becomes important. In reality, if the disc particles are close together, we expect many of the system's normal modes to become excited after even shorter times, because of large Laplace coefficients.

\begin{figure}
    \centering
    \includegraphics[width=0.48\textwidth]{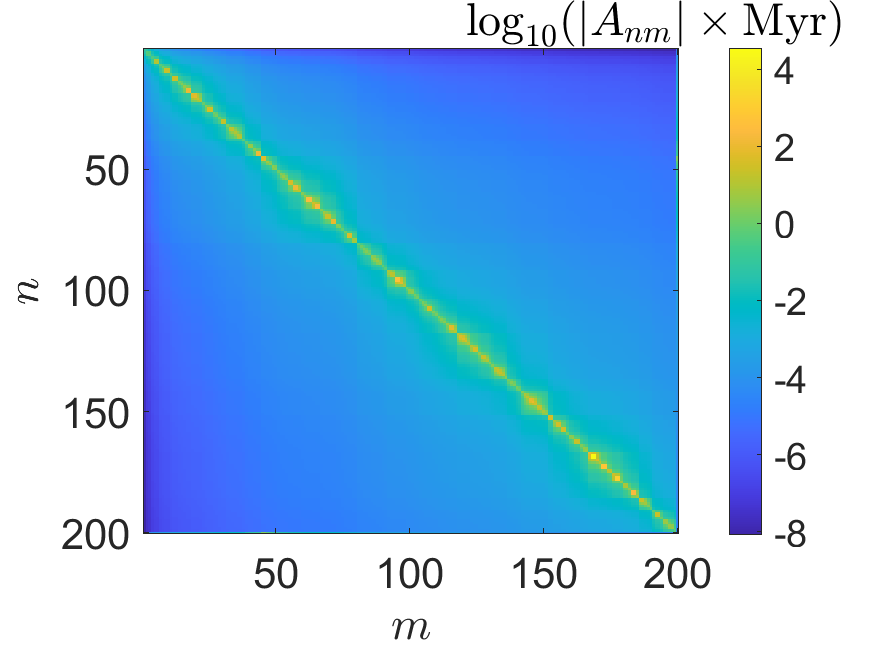}
    \caption{A plot of the matrix $A_{nm}$, in absolute value. The stars are ordered in an order of increasing semi-major axis.}
    \label{fig:a_matrix}
\end{figure}

Let us show this in the exact solution (i.e. the exact solution of the coupled harmonic oscillators). We populate a matrix $A_{nm}$ with 199 stars with $m_n = 1~M_\odot$ around a SMBH with mass $4\times 10^6~M_\odot$, and add a perturber with mass $m_{\rm p} = 20~M_\odot$, with semi-major axis $a_{\rm p} = 0.05$pc. The stars are sampled on circular orbits from a uniform distribution in $a$ between $10^{-4}$pc and $0.5$pc. The matrix $A$ that was obtained is plotted in figure \ref{fig:a_matrix}.
While the diagonal elements of $A_{nm}$ are very large, most modes that are excited by the initial conditions $\set{s_n \approx 0, 0< s_p = \sin(i_p(0)) \ll 1}$ are low frequency ones, in agreement with the approximation made above.

\begin{figure*}
    \centering
    \includegraphics[width=0.999\textwidth]{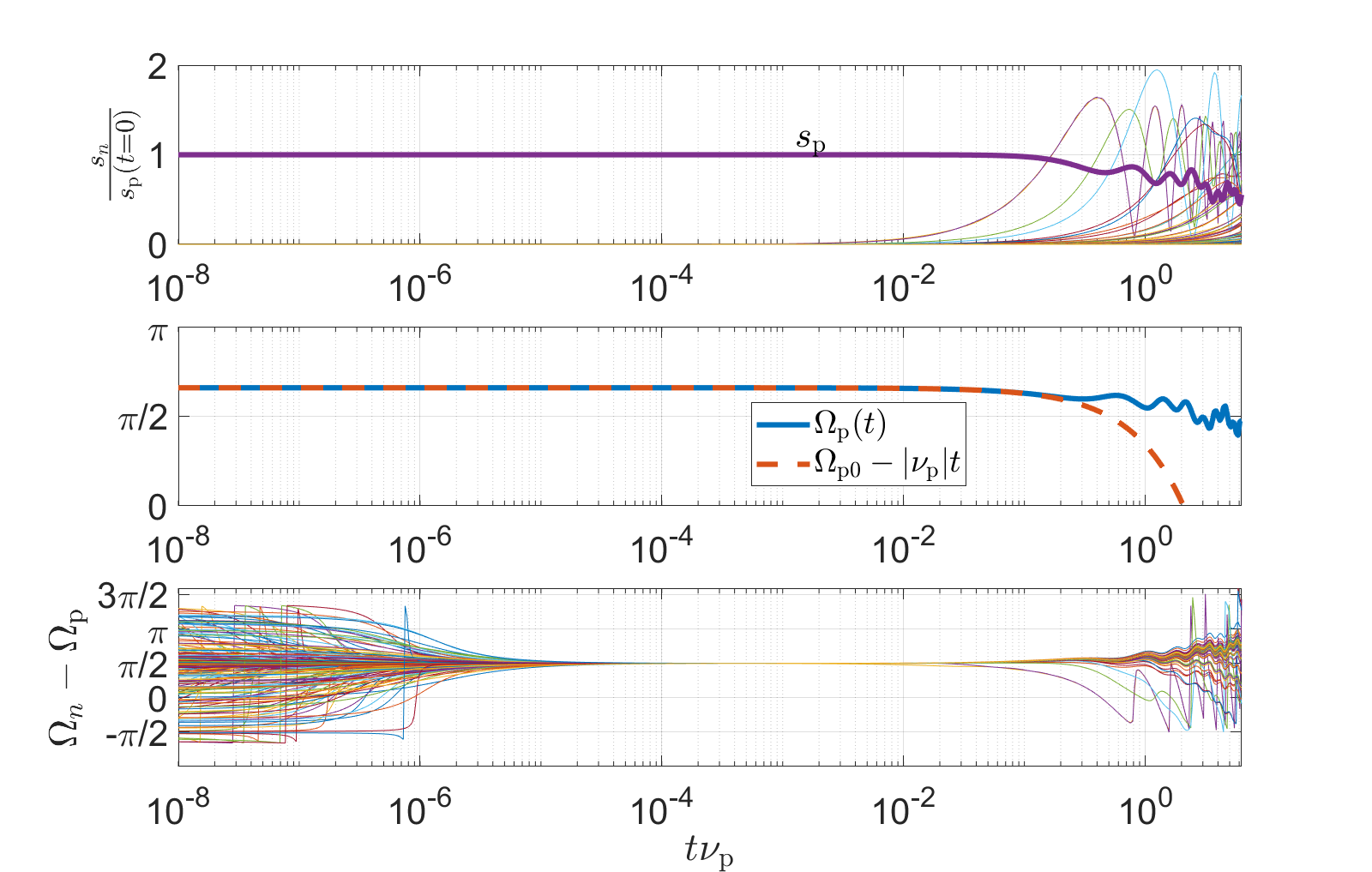}
    \caption{The inclinations (top) and arguments of the ascending node (bottom) of the disc stars, as a function of time, in units of $\nu_{p} = 2A_{pp}$. The bottom panels clearly shows an alignment of $\Omega_n$ with $\Omega_{\rm p} + \pi/2$ which occurs extremely fast, over the time-scale $b_{{\rm p}n}^{-1}$, expanded at small inclinations and eccentricities. The argument of the ascending node $\Omega_{\rm p}$ of the perturber is shown in the middle panel, with the approximation $\Omega_{\rm p0}-\abs{\nu_{\rm p}t}$, which holds until $\nu_{\rm p} t \sim 10^{-1}$. These solutions are the exact solutions to equations \eqref{eqn:equations of motion perturber Harmonic oscillator}, obtained by diagonalising the matrix $A_{nm}$.}
    \label{fig:oscillator}
\end{figure*}

One can see this in figure \ref{fig:oscillator}, which shows the exact solution to equations \eqref{eqn:equations of motion perturber Harmonic oscillator},
that the inclinations increase in an order unity time, and that at the beginning all the arguments of ascending node align extremely fast with $\Omega_n - \Omega_{\rm p} = \pi/2$. Indeed, the stars whose inclinations increase the fastest are those which deviate fastest from this alignment.
Then, at around $\nu_{\rm p}t\sim 10^{-1}$, other modes become excited with sufficiently large amplitudes (this happens earlier than at $\nu_{\rm p}t \sim 1$ because the Laplace coefficients render the diagonal elements of $A_{nm}$ very large, so that this deviation occurs when $b_{3/2}^{(1)}s_n$ ceases to be sufficiently small), and stars begin to deviate from the node alignment.
We will show below how this alignment leads to a decrease in $s_{\rm p}$, in the general case.

\section{Disc Thickening}
\label{sec: disc thickening}
After solving for the fast variables, the arguments of the ascending node, we can now proceed to solve the equations of motion for the slow variables -- namely, the inclinations. Here again we do not assume that the eccentricity or the inclination of the perturber are small. This can be done by substituting the solutions for $\Omega_n$ and $\Omega_{\rm p}$ into the equations of motion for $i_n$. Then, by angular momentum conservation, one can get $i_{\rm p}$.

One has, from equation \eqref{eqn:i n dot Lagrange equation of motion},
\begin{equation}\label{eqn:s_n dot boundary layer}
\begin{aligned}
  \frac{\mathrm{d}\cos(i_n)}{\mathrm{d}t} & \approx -A_ns_n\sqrt{[1-s_n^2]\left[1-\frac{\nu_{\rm p}^2}{b_{\textrm{p}n}^2}\right]}
  \\ &
  + \sum_{m}\frac{2A_{nm}s_ms_n \sin(\Omega_n-\Omega_m)}{\gamma_n^2},
\end{aligned}
\end{equation}
where $A_{nm}$ was introduced below Eq.~\eqref{eqn:definition p_n q_n}. This equation includes both the effect of $\ham_{\rm p}$ and that of $\ham_{\rm LL}$. Substituting $\Omega_n$ from equation \eqref{eqn: Omega_n alignment}, one finds
\begin{equation}
\begin{aligned}
  \frac{\mathrm{d}s_n}{\mathrm{d}t} & \approx \frac{3}{8}\frac{Gm_{\rm p}m_ns_{pn2}\alpha_{pn}^2}{\gamma_n^2\max\set{a_{\rm p},a_n}}\sqrt{1-s_n^2}\sqrt{1-\frac{\nu_{\rm p}^2}{b_{\textrm{p}n}^2}}\sin (2i_p) \\ &
  - \frac{2}{\gamma_n^2} \sum_{m}A_{nm}s_m \left(\frac{\nu_{\rm p}}{b_{\textrm{p}m}} - \frac{\nu_{\rm p}}{b_{\textrm{p}n}}\right).
\end{aligned}
\end{equation}
It is clear that the first term dominates for small inclinations, while for inclinations of order $\eps$, the two terms are roughly equal. We therefore expect the alignment to persist until the perturber sinks into the disc, i.e. until $\ham_{\rm LL}$ cannot be neglected relative to $\ham_{\rm p}$.
The equilibrium point, besides $s_n = 0$ for all $n$ (which is unstable) is $s_n$ such that $\nu_{\rm p} = b_{\textrm{p}n}$, i.e.
\begin{equation}
  s_n = s_{n,\textrm{late}} \equiv \frac{A_n}{\sqrt{A_n^2 + \nu_{\rm p}^2}}.
\end{equation}

The equation of motion for $J_{\rm p}^a \equiv \gamma_p^2 \cos i_{\rm p}$, the canonical conjugate of $\Omega_{\rm p}$, i.e. the $\zhat$-component of the perturber's angular momentum (where the $\zhat$-axis is defined such that initially the disc is the $\xhat$-$\yhat$ plane)
is then simply
\begin{equation}\label{eqn:early times}
  \Delta_2\dot{J}_{\rm p}^a = \sum_{n}\left[\frac{\partial^2\ham_{\rm p}}{\partial\Omega_n\partial J_n}\Delta_1J_n + \frac{\partial^2\ham_{\rm p}}{\partial\Omega_n^2}\Delta_1\Omega_n\right],
\end{equation}
where $\Delta_1$ and $\Delta_2$ pertain to the $\ord{\eps}$ and $\ord{\eps^2}$ corrections, respectively, the derivatives are evaluated along the unperturbed trajectory, as is the right-hand side of equation \eqref{eqn:s_n dot boundary layer}, and $J_n^a \equiv \gamma_n^2 \cos i_n$. This yields
\begin{equation}\label{eqn:i_n of t}
    \begin{aligned}
     & \Delta_1J_n^a = \cos(i_{n}(0))\gamma_n^2 \times \\ &
     \max\set{1-A_n(0)s_n(0)t\sqrt{1-\frac{\nu_{\rm p}^2}{b_{{\rm p}n}(0)}},\frac{\left[A_n(0)^2+\nu_{\rm p}^2\right]^{-1/2}}{\nu_{\rm p}^{-1}\cos(i_{n}(0))}}
\end{aligned}
\end{equation}
and the term involving $\Delta_1\Omega_n$ is sub-leading. The result of following through with equation \eqref{eqn:early times} is an expression for resonant dynamical friction:
\begin{equation}
    \frac{\mathrm{d}i_{\rm p}}{\mathrm{d}t} = \sum_{n: \abs{\frac{\nu_{\rm p}}{b_{\textrm{p}n}}} \leq 1} \frac{b_{\textrm{p}n}(0)\cos(2i_n(0))}{\gamma_n^{-2}\cos^2(i_n(0))} \left[\frac{\cos(i_n) - \cos(i_n(0))}{\gamma_{\rm p}^2 \sin i_{\rm p}}\right].
\end{equation}
This completes the derivation of equation \eqref{eqn:perturber's inclination equation of motion}. The dynamical friction time-scale (which is defined in equation \eqref{eqn:tau rdf definition}) is then simply
\begin{equation}
    \tau_{\rm RDF}^{-2} = \sum_{n: \abs{\nu_{\rm p}} \leq \abs{b_{\textrm{p}n}(0)}} \frac{A_n(0)^2 \gamma_n^2}{\gamma_{\rm p}^2}.
\end{equation}
For an initially thin disc, with circular orbits and a surface density $\Sigma(r) \equiv \frac{M_{\rm d}}{2\pi R^2}\sigma(r/R)$, for any function $\sigma(x)$ this becomes
\begin{equation}\label{eqn: tau rdf general surface density}
\begin{aligned}
    \tau_{\rm RDF}^{-2} & = \frac{9\pi^2\sin^2(2i_{\rm p}(0))}{16} \frac{m_{\rm p}M_{\rm d}}{M_\bullet^2} t_{\rm orb}^{-2} \\ &
    \times \left[\int_0^{a_{\rm p}} \frac{r^{9/2}}{R^2a_{\rm p}^{7/2}} \sigma\left(\frac{r}{R}\right)\mathrm{d}r + \int_{a_{\rm p}}^R \frac{a_{\rm p}^{13/2}}{R^2 r^{11/2}}\sigma\left(\frac{r}{R}\right)\mathrm{d}r\right].
\end{aligned}
\end{equation}

The power-law profile mentioned at the end of \S \ref{subsec:zeroth order} yields a dynamical friction time-scale of
\begin{equation}\label{eqn:tau RDF appendix}
\begin{aligned}
 \tau_{\rm RDF} & = \frac{4}{3\pi} t_{\rm orb}\, \frac{M_\bullet}{\sqrt{m_{\rm p}M_{\rm d}}} \frac{(3-\beta)^{-1/2}}{\sin(2i_{\rm p}(0))}\left[\frac{R}{a_{\rm p}}\right]^{1-\beta/2} \\ &
 \times \left[\frac{2}{7-2\beta} + \frac{2}{5+2\beta}\left(1-\frac{a_{\rm p}^2}{R^2}\right) \right]^{-1/2}.
\end{aligned}
\end{equation}
 For example, for $\beta = 1$ and $a_{\rm p} = R/10$, we find $\tau_{\rm RDF} =  \frac{1.15 t_{\rm orb}}{\sin(2i_{\rm p}(0))}\frac{M_\bullet}{\sqrt{m_{\rm p}M_{\rm d}}}$.

\section{Numerical Set-Up}
\label{sec:simulation}
This appendix supplements \S \ref{sec:simulation shortened}, by providing details of the numerical model.

The SMBH is modelled as an external point-mass potential. The mass of the SMBH is allowed to grow due to the tidal disruption of stars \citep{JYM2012, LLBCS2012, ZBS2014}; the tidal disruption radius sets the innermost resolution of the simulation, which we choose to be equal to twice the tidal disruption radius of the Sun by a $4\times10^6M_\odot$ SMBH. While the mutual interaction between stars is softened, with softening-length $\epsilon_\mathrm{ss}=5\times10^{-5}$pc (which prevents the formation of close binary systems), we do not soften the interaction of stars with the SMBH \citep{KCMB2018}. No relativistic effects are accounted for in the simulations, however.

To test the calculations described in the paper, we perform toy-model simulations consisting of a thin stellar disc of $N = 999$ particles (excluding the IMBH) where the angular momentum vector initial lie in a narrow cone with opening angle of $1^\circ$ (orbital inclinations are generated from a cosine-uniform distribution between $\cos{0}$ and $\cos{1^\circ}$) and assume initially nearly circular orbits with eccentricities of $e_n=0.01$. We choose the 3D stellar density profile to match the observed value for the young stellar disc in the Galactic centre, which follows a power-law density distribution $\rho\propto r^{-2.4}$ \citep{Yelda2014}. Initial longitudes of the ascending node, arguments of periapsis and mean anomalies are drawn from a uniform distribution over their entire allowed range. We generate the initial conditions assuming that particles follow Keplerian ellipses focussed at the SMBH, which is chosen to be the origin of the coordinate system. The initial inclination angle of the IMBH is 45$^\circ$ with respect to the disc plane. The entire system is embedded in a smooth Plummer potential \citep{Plummer1911}.

To make this more realistic, we also perform a set of simulations with a `live' sphere of $N_\mathrm{s} = 10^4$ with and without an analytic Plummer potential $\phi_\mathrm{Pl} = -\frac{GM_\mathrm{Pl}}{\sqrt{r^2+r_0^2}}$, where $M_\mathrm{Pl}=2\times10^5M_\odot$, $r_0=0.5$ pc; and $N_\mathrm{s} = 10^5$ particles without the external potential. We adopt the initial conditions for the live sphere and the stellar disc from \citet{Panamarev2022}: for the stellar disc, we choose the distribution of the initial orbital inclinations and eccentricities to match a disc that may have formed by interaction of stars with a gaseous accretion disc \citep{Panamarev2018}, a disc model we term `stardisc'. We also have one model where the distribution of eccentricities is thermal and inclination angles are drawn from a cosine-uniform distribution in the range between $\cos0$ and $\cos{10^\circ}$ which we label as `thermal'. The stars in the live sphere are initialised on Keplerian ellipses with eccentricities drawn from the thermal distribution, a uniform distribution of cosines of orbital inclinations, semi-major axes matching the \citet{Bahcall1976} cusp with a power-law index $\gamma=1.75$. Note, however, that a more realistic live halo would be far more massive and extend beyond the central 0.5 pc; in which case resonant relaxation processes due to the spherical halo could be important and change the evolution of the disc (see \citealt{per+18}). The evolution of such systems is not explored here, where we consider less massive systems.
Both disc and sphere are confined within the innermost region of the Galactic centre so that $\max\set{a_\mathrm{d}}=\max\set{a_\mathrm{s}} \leq 0.5$pc, where $a_\mathrm{d}$ and $a_\mathrm{s}$ are the semi-major axes of the stars in the disc and sphere, respectively. We refer to \S 3.2 of \citet{Panamarev2022} for details. We  list all the different simulations in Table \ref{tab:runs}.

\begin{table}
\centering
\caption{A list of simulation configurations used here. The models have different initial conditions for the IMBH, the spherical component, and the stellar disc. The number of stars in the disc is $N=999$ in all models, and the IMBH's semi-major axis is $a_{\rm p} = 0.05$pc.}
\label{tab:runs}
\begin{tabular}{llllll}
\hline\hline
$m_\mathrm{p}$[$M_\odot$] & $i_\mathrm{p} [^\circ]$ & $e_\mathrm{p}$ & Disc Model & $N_\mathrm{s}$ & Sphere   \\
\hline
250  & 45 & 0.33 & $e_\mathrm{d} = 0.01$ & - & Plummer \\
250  & 45 & 0.33 & stardisc & $10^4$ & live+Plummer \\
2000  & 45 & 0.9 & stardisc & $10^4$ & live \\
1000  & 45 & 0.33 & stardisc & $10^4$ & live \\
2000  & 45 & 0.33 & stardisc & $10^4$ & live \\
2000  & 45 & 0.33 & stardisc & $10^5$ & live \\
2000  & 45 & 0.33 & thermal & $10^5$ & live \\
2000  & 0 & 0.33 & stardisc & $10^4$ & live+Plummer \\
\hline
\end{tabular}
\end{table}


\bsp	
\label{lastpage}
\end{document}